\newcommand*\subtxt[1]{{\textnormal{#1}}}

\documentclass[fleqn,usenatbib]{mnras}

\usepackage{newtxtext,newtxmath}
\usepackage{xspace}
\usepackage{xcolor}

\usepackage[T1]{fontenc}

\DeclareRobustCommand{\VAN}[3]{#2}
\let\VANthebibliography\thebibliography
\def\thebibliography{\DeclareRobustCommand{\VAN}[3]{##3}\VANthebibliography}

\usepackage{graphicx}	
\usepackage{amsmath}	
\usepackage{orcidlink}
\usepackage{multirow}

\title[A Robust Analysis of QU-fitting Behavior]{A Robust Analysis of QU-fitting Behavior for 800-1088 MHz and 1296-1440 MHz}

\author[L. Oberhelman et al.]{
Lindsey Oberhelman$^{1}$\thanks{E-mail: Lindsey.Oberhelman@anu.edu.au} \orcidlink{0009-0009-2848-5269},
Cameron L. Van Eck$^{1}$ \orcidlink{0000-0002-7641-9946},
N.~M.~McClure-Griffiths,$^{1}$\orcidlink{0000-0003-2730-957X}
Yik Ki Ma$^{2,1}$
\orcidlink{0000-0003-0742-2006},
\newauthor 
Alec J. M. Thomson$^{3,4}$ \orcidlink{0000-0001-9472-041X}, 
Jason M. Price$^{1}$\orcidlink{0000-0001-7013-0562},
Shinsuke Ideguchi $^{5}$, 
Craig S. Anderson$^{1}$\orcidlink{0000-0002-6243-7879},
\newauthor 
Marijke Haverkorn$^{6}$
\orcidlink{0000-0002-5288-312X},
Denis Leahy$^{7}$
\orcidlink{0000-0002-4814-958X},
Takuya Akahori$^{8}$ \orcidlink{0000-0001-9399-5331},
Jennifer West$^{9}$ \orcidlink{0000-0001-7722-8458}\\
$^{1}$ Research School of Astronomy \& Astrophysics, The Australian National University, Canberra, ACT 2611, Australia \\
$^{2}$ Max-Planck-Institut f\"ur Radioastronomie, Auf dem H\"ugel 69, 53121 Bonn, Germany \\
$^{3}$ SKA Observatory, SKA-Low Science Operations Centre, 26 Dick Perry Avenue, Kensington WA 6151, Australia \\
$^{4}$ ATNF, CSIRO Space \& Astronomy, PO Box 1130, Bentley, WA 6102, Australia \\
$^{5}$ National Astronomical Observatory of Japan, 2-21-1 Osawa, Mitaka, Tokyo 181-8588, Japan \\
$^{6}$ Department of Astrophysics/IMAPP, Radboud University, PO Box 9010, 6500 GL Nijmegen \\
$^{7}$ Department of Physics and Astronomy, University of Calgary, Calgary, Canada T2N 1N4\\
${8}$  Mizusawa VLBI Observatory, National Astronomical Observatory Japan, 2-21-1 Osawa, Mitaka, Tokyo 181-8588, Japan \\
${9}$ Dunlap Institute for Astronomy and Astrophysics, University of Toronto, 50 St.\ George Street, Toronto, ON M5S 3H4, Canada
}

\date{Accepted 2026 February 22. Received 2026 February 11; in original form 2025 November 18}

\pubyear{\the\year{}}

\begin{document}
\label{firstpage}
\pagerange{\pageref{firstpage}--\pageref{lastpage}}
\maketitle

\begin{abstract}
QU-fitting is a powerful tool for interpreting spectro-polarimetric radio continuum observations by linking them to physical models, enabling estimates of the magnetic fields in, for example, the Milky Way, galaxy clusters, and radio jets. We present a comprehensive investigation into the effectiveness and limitations of QU-fitting within the ASKAP POSSUM survey frequency ranges (800–1088 MHz and 1296–1440 MHz) with projections to other spectro-polarimetric radio observations. We simulate different physical polarization sources: Faraday simple, Burn slab, internal turbulence, external turbulence, and two-component models in the POSSUM frequencies, and assess their observational degeneracies and fit accuracies. Our results highlight the model-dependent nature of reliable fitting and identify specific regions of parameter space where model selection, and therefore characterization of the physical medium, becomes ambiguous. For QU-fitting we find the Bayes factor, computed using the marginal likelihood, outperforms more traditionally used goodness-of-fit metrics such as Bayesian Information Criterion (BIC), Akaike Information Criterion (AIC), and $\chi^{2}$ for model selection. We provide empirical relationships to delineate the boundaries where model distinguishability is impossible. Finally, we evaluate how accurately QU-fitting recovers model parameters and their associated uncertainties, thereby assessing its ability to correctly characterize the Faraday-rotating medium in both point and extended sources in Faraday depth space.
\end{abstract}




\section{Introduction}
 
Magnetic fields in astrophysical environments are commonly studied using radio continuum observations of synchrotron emission, which is intrinsically linearly polarized. The Stokes parameters \textit{I}, \textit{Q}, and \textit{U} describe the linear polarization of the emission, and together they form the complex polarization (\textbf{P}) as:
\begin{equation}
     \textbf{P} = Q + iU = p I {\rm e}^{2i\psi},
\end{equation}
where $\psi$ is the observed polarization angle and $p$ is the measured magnitude of  linear polarization (measured fractional polarization):
\begin{equation}
    p = \sqrt{\left( \frac{Q}{I}\right)^{2} + \left( \frac{U}{I}\right)^{2}}.
\end{equation} By analysing the variation of \textbf{P} across frequency, it is possible to infer the characteristics of the intervening magnetic fields along the line of sight.

As synchrotron radiation propagates through a magneto-ionic medium containing free electrons and magnetic fields, its polarization angle is rotated due to the effect known as birefringence, a phenomenon known as Faraday rotation \citep[see][for a review]{Ferri2021}. The magnitude of this effect is proportional to the line-of-sight magnetic field strength, path length, and the integrated electron density; and increases with the square of the wavelength. This relationship can be expressed through the Faraday depth ($\phi$), 

\begin{equation}
    \phi \:  \mathrm{[rad \: m^{-2}]} = \frac{e^{3}}{2 \pi m_{e}^{2}c^{4}} \int_{0}^{r} n_{e} B_{\parallel} dr^{'}
\end{equation}
where $B_{\parallel}$ [$\mu G$] is the magnetic field along the line of sight (LOS), $n_{e}$ [cm$^{-3}$] is the thermal electron density, $e$ is the charge of an electron, $m_{e}$ is the mass of an electron, $c$ is the speed of light, and $r$ is the path length. In the simplest case, the synchrotron emission from the emitting volume passes through a single foreground magnetic field with homogeneous Faraday depth and experiences a Faraday rotation of some amplitude or Rotation Measure (RM). Any other case that deviates from this model will add complexity to the observed Stokes $Q$ and $U$ continuum. 

The QU-fitting technique was developed to be able to measure Faraday depth, quantify depolarization, and identify complex magneto-ionic structures of astrophysical sources \citep{Miyashita2019, Skilling2004, OSullivan2012, Farnsworth_2011, Law2011}. Since the 1960s, several models have been derived to encompass different scenarios involving uniform and turbulent magnetic fields (see \citealt{Burn1967, Sokoloff1988, Tribble1991} for derivations).
These models try to account for depolarization effects internal and external to the source. These environments include magnetized extragalactic radio source environments \citep{OSullivan2012, OSullivan2017, Pasetto2018}, probing the magnetized interiors of radio lobes \citep{Kaczmarek2018, Anderson2018, Andati2024}, and galactic foreground \citep{Ranchod2024, Thomson2021}.

QU-fitting is hindered by two kinds of degeneracy. (1) Mathematical degeneracy, occurring when distinct physical models are identical when convolved by a telescope beam or in the limit where a parameter reaches an extreme value. This is demonstrated in \cite{Sokoloff1988} and in Appendix \ref{app:Degen_Proof}. (2) Observational degeneracy, occurring under conditions imposed by a finite bandwidth, within which two different model spectrum may be indistinguishable in parts of their parameter space. This problem becomes more acute as a model approaches a mathematically degenerate limit with another model, but its exact extent depends on the observational bandwidth and must be assessed for each case individually. 


There have been prior studies that have aimed to understand these model observational degeneracy problems by simulating QU-spectra for different line of sight models and testing how QU-fitting handles these cases. \citet{Sun2015} was the first to systematically study QU-fitting algorithms. This study was an investigation performed in preparation for the Polarization Sky Survey of the Universe's Magnetism (POSSUM; \citealt{Gaensler2010}, \citealt{Gaensler2025}) survey to test the effectiveness of methods like QU-fitting. They tested different source types (i.e.  Faraday thick and thin, single component, and two components) in the frequency range 1100-1400 MHz.  Their results indicated that QU-fitting is inaccurate when there is more than one component but more accurate than other algorithms. They predicted that users will encounter degeneracy challenges with thick models, though it is hard to pinpoint exactly where these challenges will occur in model parameter space. 

\citet{Miyashita2019} simulated two sources along the LOS and showed that, for the two-component model, QU-fitting accuracy will decrease when the components are close enough together in Faraday depth space. They used the Akaike Information Criterion (AIC), Bayesian Information Criterion (BIC), and $\chi^{2}$ for model fitness, finding that $\chi^{2}$ is a poor indicator of successful convergence. The findings of  \citet{Miyashita2019} and  \citet{Sun2015} underscore the need for further work to specify and define the pitfalls of QU-fitting techniques and address challenges with model degeneracy.

There are many new large-scale radio continuum surveys with wide bandwidths and high spectral resolution, such as POSSUM, the MeerKAT International GHz Tiered Extragalactic Exploration (MIGHTEE, \citealt{Jarvis2016}), and the Global Magneto-Ionic Medium Survey (GMIMS, \citealt{Ordog2025}, \citealt{Wolleben2009}). However, they will be subject to all the aforementioned degeneracy challenges with QU-fitting, and ambiguity surrounding the accuracy of the fitted parameters. 

Here we investigate the best metrics for QU model selection and parameter estimation accuracy for the POSSUM survey focusing specifically on the low-band capabilities but also considering the full-band capabilities that will likely come online in a few years. We also make estimates of the potential limitations for other similar surveys. To meet these goals, we conducted a series of QU-fitting `experiments' with different models. In this work, we simulated both Faraday `thin' or `screen' models, as well as Faraday `thick' or `slab' models. We also tested two-component models and different small-scale turbulence models.

This paper is structured as follows. We first present the models that we test in Section \ref{Models}. Then, we describe how we set up our simulations (Section \ref{Methods}) and how we conduct our QU-fitting. We discuss our results of the observational degeneracy experiments of the individual model (Section \ref{Models_Exp_Results}), and the precision of the fitted parameters and the estimation of the QU-fitting error in Section \ref{Accuracy}. In Section \ref{Discussion} we present our best use guidelines for QU-fitting with the POSSUM survey and make predictions for other surveys. 

\section{Models}\label{Models}

In this Section we define and motivate the theoretical models that we tested and lay out the different experiments we performed. Many models for line of sight physics have been developed, see \cite{Burn1967}, \cite{Sokoloff1988}, and \cite{Tribble1991} for full derivations. We selected five diverse and frequently used models that are building blocks for more complex models to be built from, like those developed by \citep{Ma2025, OSullivan2012}. 

\subsection{Model descriptions}
\subsubsection*{Faraday simple model (FS)}
The most basic model is the Faraday simple model or ``thin model" (FS), 

\begin{equation}
    \textbf{P}(\lambda^{2}) = p_{0}{\rm e}^{2i(\psi_{0}+\text{RM}_{\subtxt{screen}} \lambda^{2})},
    \label{FS}
\end{equation}

where $p_{0}$ is the intrinsic fractional polarization of the emission, $\psi_{0}$ is the intrinsic polarization angle, and $\text{RM}_{\subtxt{screen}}$ is the rotation measure of a foreground screen. This model is comprised of a synchrotron-emitting source and a separate magnetized thermal plasma that performs the Faraday rotation. A $\text{RM}_{\subtxt{screen}}$ term is always present in all of our models because we assume that there is always some amount of foreground Faraday rotation on top of any internal or differential Faraday rotation. 
\vspace{-0.5cm}
\subsubsection*{Two-component model (TC)}
The two distinct Faraday simple components or `two-component model' (TC) is modeled as
\begin{multline}
   \textbf{P}(\lambda^{2}) = p_{0,1}{\rm e}^{2i(\psi_{0,1}+\text{RM}_{\subtxt{screen, 1}}\lambda^{2})} + \\ p_{0,2}{\rm e}^{2i(\psi_{0,2}+\text{RM}_{\subtxt{screen, 2}}\lambda^{2})}, 
   \label{TC} 
\end{multline}
where $\text{RM}_{\subtxt{screen, 1}}$ and $\text{RM}_{\subtxt{screen, 2}}$ each represent a distinct component along the LOS. As the name indicates, this model describes a scenario in which the line of sight contains two separate emission components, each experiencing different foreground Faraday rotation—effectively two Faraday simple models. It can approximate both extragalactic unresolved sources and resolved sources viewed through the Milky Way. While the model is commonly extended to include more components, we restrict our analysis to two in this study. 

\vspace{-0.5cm}
\subsubsection*{Burn slab model (BS)}
In contrast to the thin models, `slab' models represent the scenario where the synchrotron emission volume is within a Faraday rotating plasma. The most basic of these models is the Burn slab, which assumes that the uniform thermal plasma within the source is magnetized with a large-scale homogeneous magnetic field. This leads to a differential Faraday rotation or “Faraday width" in the source. The emission originating on the far side of the slab will experience a different rotation than the front. When summed along one line-of-sight this leads to depolarization that scales with the Faraday width ($\text{RM}_{\subtxt{src}})$ as:
\begin{equation}
    \textbf{P}(\lambda^{2}) = p_{0}\frac{\sin(\text{RM}_{\subtxt{src}}\lambda^2)}{\text{RM}_{\subtxt{src}}\lambda^2}{\rm e}^{2i(\psi_{0}+(\frac{1}{2}\text{RM}_{\subtxt{src}}+{\text{RM}_{\subtxt{screen}})\lambda^{2})}}.
    \label{BS}
\end{equation}
Depending on the source, Burn slabs are used for both resolved and unresolved radio galaxies.
\vspace{-0.5cm}
\subsubsection*{Internal turbulence model (In-Turb})
The previous three models describe only the effects of the regular, large-scale magnetic field. There are often small-scale magnetic fields that need to be considered. The internal turbulence model describes the slab model, including the turbulent field of the thermal plasma. In this case, the slab is represented as: 
\begin{equation}
    \textbf{P}(\lambda^{2}) = p_{0} \frac{1-{\rm e}^{2i\lambda^{2}\text{RM}_{\subtxt{src}} - 2\lambda^{4}\sigma_{\subtxt{RM,src}}^{2}}}{2\lambda^{4}\sigma_{\subtxt{RM,src}}^{2} - 2i\lambda^{2}\text{RM}_{\subtxt{src}}} 
{\rm e}^{(2i(\psi_{0} + \text{RM}_{\subtxt{screen}}\lambda^{2}))}.
\label{T-slab}
\end{equation}

The $\sigma_{\subtxt{RM,src}}$ parameter represents the standard deviation of the Faraday rotation  caused by the small-scale magnetic field. This model represents the physics of turbulence within the source's magnetic field. 
\vspace{-0.5cm}

\subsubsection*{External turbulence model (Ex-Turb})
The external turbulence model (Ex-Turb) describes an emission source that has no internal Faraday rotation with a foreground thermal plasma with a small-scale magnetic field:
\begin{equation}
    \textbf{P}(\lambda^{2}) = p_{0} {\rm e}^{2i(\psi_{0}+\text{RM}_{\subtxt{screen}}\lambda^{2})}  {\rm e}^{-2\lambda^{4}\sigma_{\subtxt{RM, FG}}^{2}}. 
    \label{T-Screen}
\end{equation}
Where $\sigma_{\subtxt{RM, FG}}$ is the turbulent field component of a foreground plasma. These two turbulence models are mostly used to model the structure of resolved and unresolved radio lobes \citep{Andati2024, Kaczmarek2018}. 


\subsection{QU-fitting experiments with model populations}\label{Model_Experiments}
We conducted four model experiments to investigate and quantify the observational degeneracy limitations of QU-fitting. These experiments were conducted on simulated populations of each of the four models which were fit with both their true model and an observationally degenerate alternative model. The experiments are as follows:

\begin{enumerate}
  \item Faraday simple and Burn slab.
  
        These two models become mathematically degenerate in the limit of $\text{RM}_{\subtxt{src}}\rightarrow 0 $rad m$^{-2}$. We expect QU-fitting to reach an observational degeneracy limit before then, at small slab widths when the algorithm can no longer distinguish between the two models. This effect depends on survey bandwidth and is not well quantified in previous literature. 
        
  \item Burn slab and two-component.
  
        This experiment primarily aims to test both how well QU-fitting can separate the spectra of one thick component from two simple components, but has implications for selecting between FS and TC as well. The work of \cite{VanEck2017} indicates that these models may be degenerate in low frequency ranges.
        
  \item Burn slab and internal turbulence.
  
        The internal turbulence model has two parameters of interest ($\text{RM}_{\subtxt{src}}$ and $\sigma_{\subtxt{RM,src}}$) making it more complex than the Burn slab. These two models are mathematically degenerate in the limit as $\sigma_{RM} \rightarrow 0$ rad m$^{-2}$, and it can be difficult to disentangle the depolarization effects of $\text{RM}_{\subtxt{src}}$ and $\sigma_{\subtxt{RM,src}}$ in Faraday depth space as seen in Figure \ref{dict_BS_TS}. 
        
  \item Internal turbulence and external turbulence.
  
        These two turbulence models are commonly used for resolved radio lobes. The model selection for these two models is affected by both the regular magnetic field component $\text{RM}_{\subtxt{src}}$ (just like the Burn slab) and the turbulent field component ($\sigma_{\subtxt{RM, src}}$ and $\sigma_{\subtxt{RM, FG}}$). We expect that, while not sharing the exact same degeneracy relationship as the slab and screen, there are sections of parameter space where the added small-scale turbulence parameter makes selection easier and where it makes it more difficult.
        
\end{enumerate}
In addition to these experiments, we also provide an analysis of the accuracy of the fitted parameters and the quality of the measurement uncertainties produced by QU-fitting. 

\section{Experimental Methods}\label{Methods}
To ensure reliable and minimally biased results, we generated populations of 10,000 fractional $q$ and $u$ (where $q=\frac{Q}{I}$ and $u=\frac{U}{I}$) spectra for each model with noise added to each spectrum. We constructed these populations for both the POSSUM low-band (800–1088 MHz) and the full-band, which combines the low-band with the mid-band (1296–1440 MHz), using 1 MHz channelization \citep{Gaensler2025}.

Model parameters were drawn from uniform distributions (Table \ref{tab:parameter_dist}) spanning ranges reported in the literature for both unresolved and resolved radio sources \citep{VanEck2018}. 
We limit the maximum range of fractional polarization values to 0.5 to be more reflective of real source populations \citep{Rudnick2014}. The upper limits on $\text{RM}_{\subtxt{src}}$, $\sigma_{\subtxt{RM,src}}$, and $\sigma_{\subtxt{RM,FG}}$ are set by the POSSUM band-averaged fractional polarization for each model. We created the maximum limit for these parameters so that our source are above the POSSUM on-axis leakage–corrected detection limit of $< 1 \%$ \citep{Gaensler2025}. For the Burn slab model, the polarized emission remains above this limit up to $\text{RM}_{\subtxt{src}}$ = 100\ $\mathrm{rad\ m^{-2}}$, so we adopt this as the upper bound. For the In-Turb model, full depolarization occurs at $\sigma_{\subtxt{RM,src}} \gtrsim 60\ \mathrm{rad\ m^{-2}}$ when $\text{RM}_{\subtxt{src}} = 100\ \mathrm{rad\ m^{-2}}$, so we set 60 $\ \mathrm{rad\ m^{-2}}$ as the limit. For the Ex-Turb, we set $\sigma_{\subtxt{RM,FG}}$ = 25 $\mathrm{rad\ m^{-2}}$, above which sources are fully depolarized in the POSSUM bands. 

\begin{table}
	\centering
	\caption{The range of uniform distributions from which the model parameters are drawn (left) and the priors we adopted for the Bayesian inference (right).}
	\label{tab:parameter_dist}
	\begin{tabular}{lccr} 
		\hline
		Model Parameter & Distribution Range & Prior Range \\
		\hline
		$p_{0}$ & 0 - 0.5  & 0 - 0.8\\
		$\psi_{0}$ & 0\textdegree - 180\textdegree & 0\textdegree - 180\textdegree\\
		  $RM_{\subtxt{screen}}$ & $-$1100 - +1100 rad m$^{-2}$ & $-$1100 - +1100 rad m$^{-2}$\\
        $RM_{\subtxt{src}}$ & 0 - 100 rad m$^{-2}$ & 0 - 100 rad m$^{-2}$\\
        $\sigma_{\subtxt{RM, src}}$ & 0 - 60 rad m$^{-2}$ & 0 - 80 rad m$^{-2}$\\
        $\sigma_{\subtxt{RM, FG}}$ & 0 - 25 rad m$^{-2}$ & 0 - 35 rad m$^{-2}$\\
        S:N & 3 - 50\\
		\hline
	\end{tabular}
\end{table}

The final step in generating the source population was the addition of noise to the simulated complex polarization spectra. Because a primary aim of these experiments was to assess how the band-averaged signal-to-noise ratio (S:N) influences model accuracy and degeneracy, we treated the band-integrated S:N as an explicit simulation parameter. For each simulated source, we drew a target band-averaged S:N in linear polarization from a uniform distribution between 3 and 50. A noise realization was drawn separately for Stokes \textit{Q} and \textit{U} for every channel from a normal distribution with variance such that it would achieve the desired S:N ratio (see Appendix~\ref{SN}) for each frequency channel. Noise realizations were drawn independently for $q$ and $u$ in each channel. This procedure ensures that the noise is frequency-independent and uncorrelated between channels and Stokes parameters. The low-band and full-band source populations were generated independently using the same procedure.

Once the populations were created we fitted them with both the true model and other alternative models following the experiments described in Section \ref{Model_Experiments}. For each model a population of low-band sources and full-band sources was tested. For all models, parameter estimation and model selection were performed using Bayesian nested sampling. Unless otherwise stated, the priors used during fitting matched the distributions used to generate the simulated sources (Table \ref{tab:parameter_dist}).

We used the QU-fitting module from \texttt{RM-Tools}~\citep{VanEck2026}, to perform the fitting.
\texttt{RM-Tools} uses the \texttt{bilby} library \citep{bilby_paper} which includes nested sampling methods as well as MCMC methods to explore the hyperparameter space of each model, selecting by maximization of the likelihood \citep{Skilling2004}. We performed a basic performance test between the models, comparing the quality of the fits between \texttt{dynesty}, \texttt{pymultinest}, and \texttt{nestle}. We found  that \texttt{dynesty} is more consistent for the higher dimensional models \citep{S2019}. Additional work by \cite{VanEck2026}, \cite{Samajdar2022}, and  \cite{Albert2024} supports our decision. The \texttt{dynesty} sampler is a Python implementation of the dynamic nested sampling algorithm \citep{S2019, Higson2019}. Dynamic nested sampling functions are used to provide more accurate evidence estimates and larger numbers of independent posterior samples.

The outputs of QU-fitting are the best-fit parameters for the model, the error estimates on those model parameters, and goodness of fit measurements: AIC, $\chi^{2}$, BIC, and Bayesian evidence (ln(EVD)). In this analysis, we compare these metrics to establish the metric best suited for model selection. The BIC and AIC are calculated from the minimum $\chi^{2}$ value, which corresponds to the maximum likelihood estimate under the assumption of normally distributed errors. 
The BIC is defined as:
\begin{equation}
    \text{BIC} = -2\ln(\textit{L}) + k\ln(n),
\end{equation}
where \textit{L} is the maximum likelihood of the model, \textit{n} is number of data points, and \textit{k} is the number of parameters \citep{schwarz1978bic}. The AIC is defined as:
\begin{equation}
    \text{AIC} = -2\ln(\textit{L}) + 2k,
\end{equation}
\citep{akaike1974aic}.
The main difference between these is that the BIC imposes a stronger penalty on complexity that increases with sample size. The BIC is the most commonly used goodness-of-fit metric for model selection in QU-fitting \citep{Schnitzeler2018, OSullivan2012, Skilling2004}. 

 Bayesian evidence, or marginal likelihood \citep{Westerkamp2024}, measures how well a model explains the data, averaged over all possible parameter values. It requires evaluating a high-dimensional integral in the parameter space of the model, which is computationally expensive. Nested sampling makes this feasible by transforming the problem into a one-dimensional integral, allowing efficient exploration of the most relevant regions of parameter space and enabling accurate evidence estimation with fewer computations. Bayesian evidence has generally not been used for the selection of QU-fitting models, with the exception of a few studies \citep{Thomson2021}. Bayesian evidence offers a potential rigorous approach to model comparison through the calculation of the Bayes factor (BF), which quantifies the relative probability of two competing models. For model selection, we use the natural logarithm of the Bayes factor between two models, Model 1 and Model 2, calculated as: 
\begin{multline}
    \ln(\text{BF})_{\text{True Model, Alternate Model}} =  \\
    \ln(\text{EVD}_{\text{True Model}}) - \ln(\text{EVD}_{\text{Alternate Model}}).
    \label{Bayefactor}
\end{multline}

From the four possible fit metrics, we conducted a series of tests summarized in Appendix \ref{Computation} to establish if there is one metric that is superior to the rest. We conclude that the Bayes factor is the only metric that can correctly classify all the model populations. We used the Bayes factor for all of the following experiments. In Table \ref{tab:binning} we show our confidence strength scale of the Bayes factor established by \cite{jeffreys1961}, that we provide additional empirical support for in  Appendix \ref{TPR}.

We quantified numerical uncertainty (\citealt{Sivia2006}) in the Bayesian evidence by computing mean error on log(EVD) for each model: 0.38 (FS), 0.39 (BS), 0.48 (TC), 0.17 (Ex-Turb), and 0.40 (In-Turb). These uncertainties affect only marginal model comparisons near our selection strength thresholds (Table \ref{tab:binning}). Consequently, we report only strong Bayes factors certainties to ensure robust conclusions.

\begin{table}
    \centering
    \caption{Confidence strength scale for the Bayes factor as defined and motivated in \citet{jeffreys1961}.}
    \label{tab:binning}
    \begin{tabular}{lc}
        \hline
        Category & $\ln(\mathrm{BF})$ Range \\
        \hline
        Incorrect Model Preferred & $< 0$ \\
        Inconclusive & $0 \leq \ln(\mathrm{BF}) < 1$ \\
        Weak Support & $1 \leq \ln(\mathrm{BF}) < 2$ \\
        Moderate Support & $2 \leq \ln(\mathrm{BF}) < 5$ \\
        Strong Support & $\geq 5$ \\
        \hline
    \end{tabular}
\end{table}

\section{Results}
We examine observational degeneracies that limit model distinguishability, then the accuracy of QU-fitting parameter estimation across all models. We split the analysis up into low-band and full-band and provide only low-band results in the main text, while the full-band parameter maps are in the Appendix \ref{app:AccuracyPlots}. In Appendix \ref{app:Atlas}, we provide an atlas of QU spectra for each of the models that is useful for discussion of observational degeneracy.

\begin{figure*}[P]
	\includegraphics[width=0.65\textwidth]{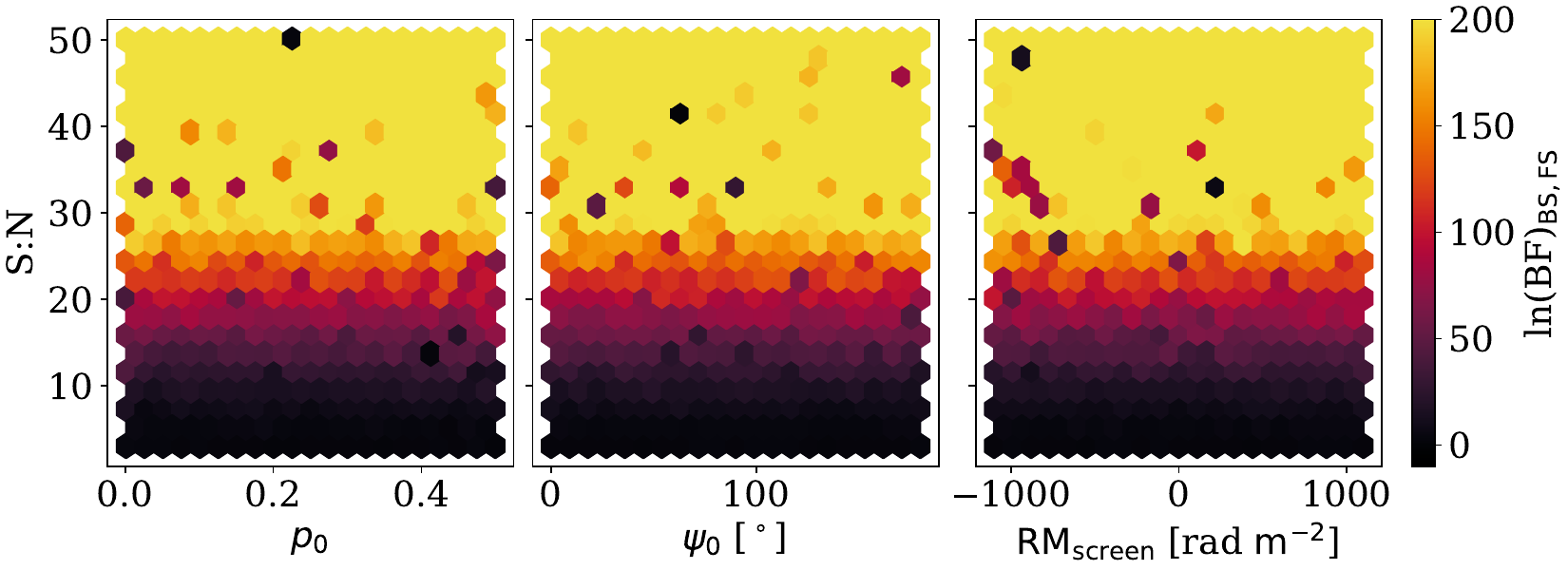}
    \caption{The parameter selection maps for the POSSUM low-band (800-1088 MHz) Burn slab population when fit with the Faraday simple model as the alternative model. These three maps are defined by S:N (y-axis) and three of the four model parameters (x-axes). From left to right they are fractional polarization, initial polarization angle, rotation measure of the uniform foreground. Each bin is colored by the median ln(BF) of the sources within ($\sim$ 25 sources per bin). See the text for explanation of the high S:N outlier bins.}
    \label{fig:allparams}
\end{figure*}

\subsection{Model Selection Degeneracy}\label{Models_Exp_Results}
The four experiments we performed are designed to compare models that are expected to be observationally degenerate within certain regions of the model parameter space. Where the degeneracy occurs depends on the physics being presented in the model and the survey bandwidth. The following sections will define the regions of model parameter space where correct model selection is possible for each of the four experiments. 

We start each experiment analysis by isolating the parameters that are the most influential for correct model selection. Then we create 2D-histograms of the space defined by these parameters with bins colored by the median (computed over all sources within each hexagon-shaped bin) Bayes factor between the two models in the experiment (Equation \ref{Bayefactor}). We refer to these histograms as `parameter selection maps'. We also create the `true positive rate maps', that are colored by the percentage of the bin population with ln(BF) > 2.0 (a correct selection with strong support). For the purpose of validating our strength scale is in alignment with probability we look at validity of our selection strength criteria with `true positive rate maps'. 

\subsubsection{Faraday simple and Burn slab}

\begin{figure*}
	\includegraphics[width=0.65\textwidth]{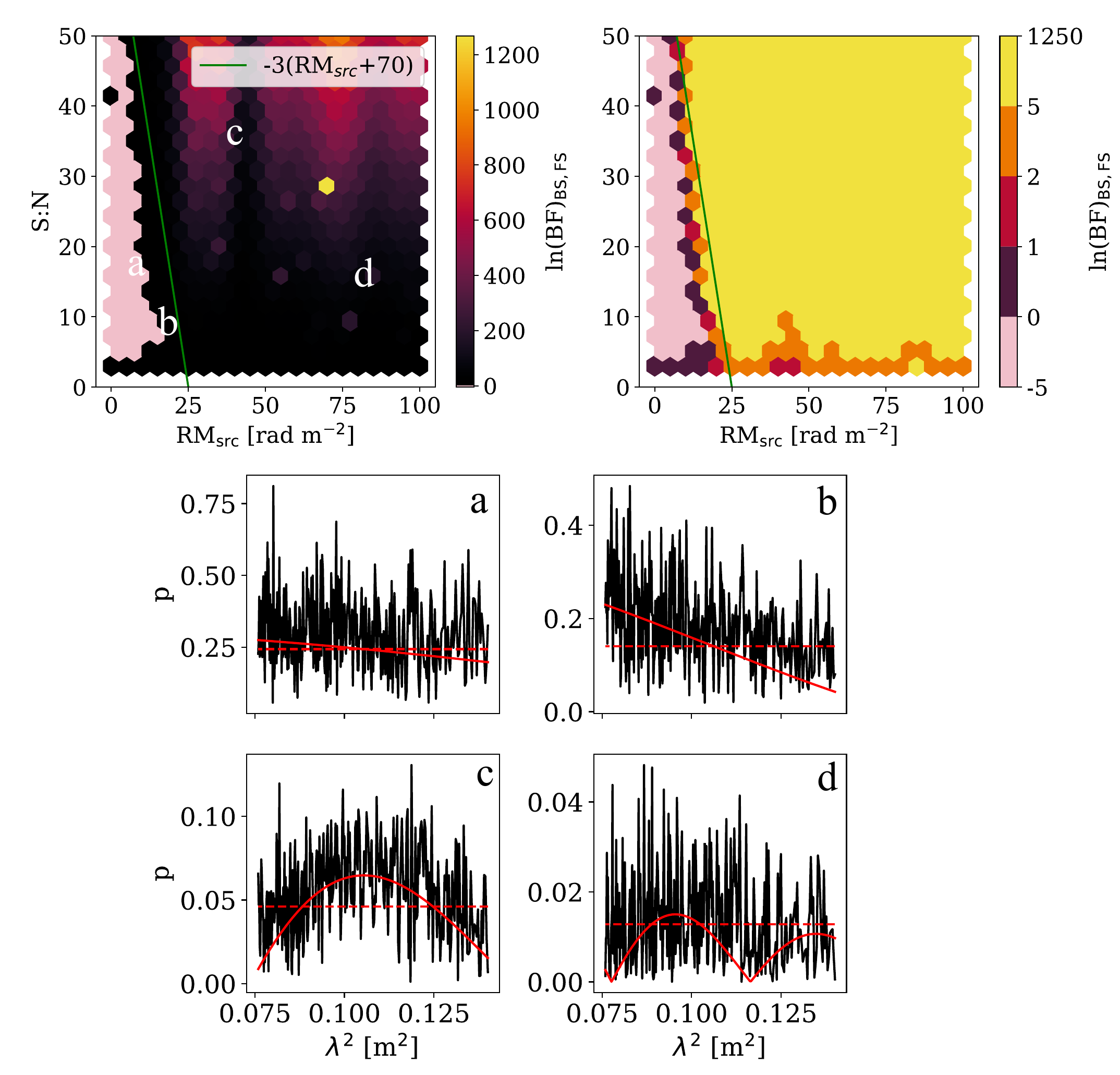}
    \caption{\textit{Top:} The parameter selection maps are shown for the POSSUM low-band (800-1088 MHz) Burn slab population fitted with both the Burn slab and Faraday simple models. Both maps are defined by the S:N and the $\text{RM}_{\subtxt{src}}$. Each bin is colored by the median ln(BF) of the sources within ($\sim$ 25 sources per bin). \textit{Top Left:} Bins are coloured according to the raw median values. \textit{Top Right:} The same quantity is shown, but using a colour scale that reflects the relative confidence in each bin (Table~\ref{tab:binning}). The larger the value, the higher the confidence that the sources in that bin are Burn slabs. The green line defines the boundary between confident correct and incorrect selections (Equation \ref{FSBSrelation}).  \textit{Bottom:} The four spectra (black) are sources taken from the corresponding lettered positions in the parameter space. The red solid lines show the Burn slab fit and the red dashed lines show the simple fit. Spectrum [a] is a thin slab that QU-fitting prefers to fit as FS, spectrum [b] is a source where the BS model is preferred, spectra [c] and [d] are sources that are in regions where the nulls of the spectra have aligned with the frequency range and so QU-fitting has elevated confidence in the FS fit.}
    \label{fig:BF_parameter_space}
\end{figure*}

The first experiment compares the Faraday simple (Equation \ref{FS}) and Burn slab (Equation \ref{BS}) models to define the limit of observational degeneracy for POSSUM bandwidths in the model parameter space. We begin with the true Burn slab population, characterized by five parameters: $p_{0}$, $\psi_{0}$, $\text{RM}_{\subtxt{screen}}$, $\text{RM}_{\subtxt{src}}$, and S:N. To reduce the dimensionality of our investigation, we first present selection maps for the parameters $p_{0}$, $\psi_{0}$, and $\text{RM}_{\subtxt{screen}}$ in Figure \ref{fig:allparams}. 
We observe only a selection dependence on S:N and not on $p_{0}$, $\psi_{0}$, or $\text{RM}_{\subtxt{screen}}$, for the remaining model selection analysis we collapse along these axes. The high S:N, low ln(BF) outliers are due to some bins randomly having higher populations of outliers, but are otherwise consistent with the statistics of the overall population. 
We now examine the relationship between S:N, $\text{RM}_{\subtxt{src}}$, and correct model selectability. In Figure \ref{fig:BF_parameter_space} (top row) we show the parameter selection maps of the S:N -- $\text{RM}_{\subtxt{src}}$ space. For both the low-band and full-band S:N > 6 is necessary for strong, confident selections in this experiment. QU-fitting confidently classifies slabs above $\text{RM}_{\subtxt{src}} \approx$ 12 – $25$ rad m$^{-2}$, depending on S:N, following:
\begin{equation}
    \text{S:N} > -2.8\ \text{RM}_{\subtxt{src}} + 70,
    \label{FSBSrelation}
\end{equation}
for the low-band only population, and
\begin{equation}
    \text{S:N}> -3.8\ \text{RM}_{\subtxt{src}} + 70
    \label{FSBSrelationcomb}
\end{equation}
for the full-band population, as shown in Figure \ref{fig:combined_band_selection}a. These empirical relations mark the boundary where the Bayes factor transitions from inconclusive to weak/moderate (from ln(BF) $\leq$ 1 to ln(BF) $\geq$ 1) support for the BS model for 100\% of the sources per bin. 

We also observe wave-like patterns in Figure \ref{fig:BF_parameter_space}, top left panel, with dips in selection confidence near $\text{RM}_{\subtxt{src}} \approx$ 45 and 85 rad m$^{-2}$. These features arise when the nulls (zeroes) of the BS spectra (sinc function) align with the band edges, making the Burn slab well fit by the FS model, even at high S:N (see spectra c and d in Figure \ref{fig:BF_parameter_space}). Using the full-band frequencies reduces this effect, so even broader bandwidths would likely mitigate it further. 


For the true Faraday simple population, a similar investigation shows that the simple model is correctly selected for 100 \% of the sources as long as S:N $\geq$ 7. Under this limit QU-fitting fits thin slabs. We find that the POSSUM full-band does not significantly improve the S:N limit for this model. 

\subsubsection{Burn slab and two-component}\label{BS_TC}
The second experiment compares the Burn slab (Equation \ref{BS}) and the two-component models (Equation \ref{TC}). We begin with model selection outcomes for a population of true Burn slabs, fitted with the TC model as the alternative. As before, we find that the significant parameters are S:N and $\text{RM}_{\subtxt{src}}$. The results for both the low-band and full-band (Figure \ref{fig:BS_TC_fullselection}) are similar to those of the previous experiment. At small slab widths, QU-fitting tends to model the BS spectra with the TC model featuring small component separations (see the first column of Figure \ref{dict_BS_MS} and the second column of Figure \ref{dict_RFG}), with the full-band data providing a slight improvement. The separation of slabs from screens in the ultra–thin slab regime will require frequencies lower than those available with POSSUM, in order to capture more of the BS sinc function. From the individual spectra shown in Figure \ref{fig:BS_TC_fullselection}, we see that the regions of reduced confidence in the full parameter selection maps correspond to the same phenomenon discussed in the previous section.

\begin{figure}
    \centering
    \includegraphics[width=0.9\linewidth]{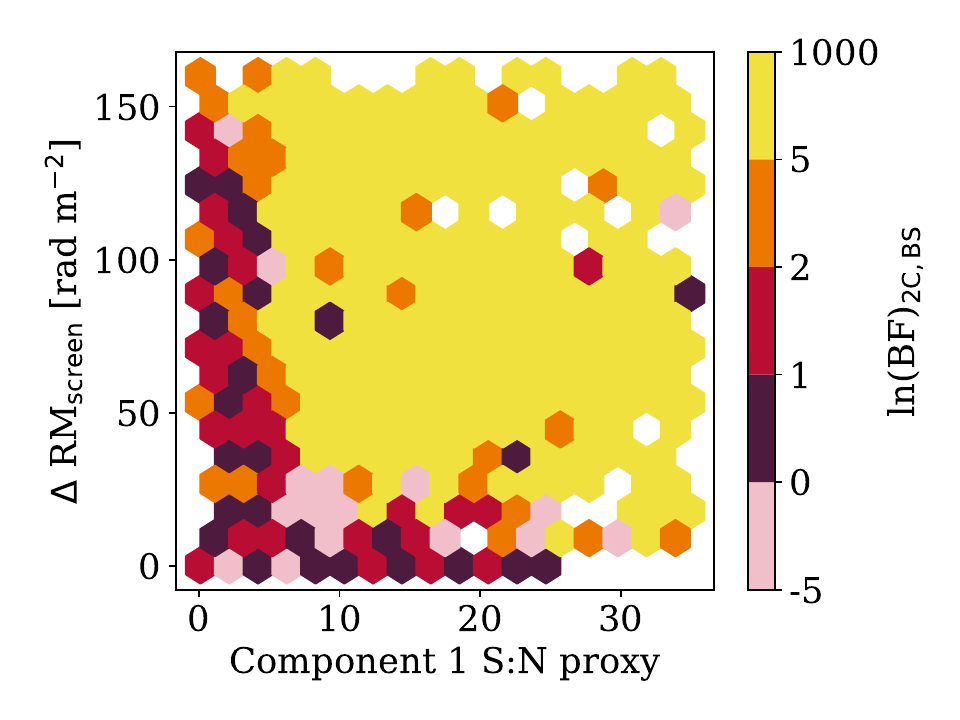}
    \caption{The parameter selection map for the POSSUM low-band (800-1088 MHz) two-component source population when the Burn slab model is the alternate. This map is defined by an estimated signal-to-noise for the first component and $\Delta \text{RM}_{\subtxt{screen}}$. We are marginalizing over the S:N for the second component, only including sources that have S:N > 5 for the second component. Each bin is colored by the median ln(BF) value and the color bar is scaled according to Table \ref{tab:binning}.}
    \label{fig:BSTCproxy}
\end{figure} 

We next turn our attention to the population of true two-component sources when the BS model is the alternate. We found no correlation between selection outcomes and the model component parameter ($\text{RM}_{\subtxt{screen}}$, $p_{0}$, or $\psi_{0}$) values. We explored the component parameter separations and found that there is no correlation between $\Delta \psi_{0}$ and confident model selection; however, there is a threshold in $\Delta \text{RM}_{\subtxt{screen}} =  \text{RM}_{\text{screen, 1}}-  \text{RM}_{\text{screen, 2}}$, where correct selection becomes difficult when the separation drops below approximately 100 rad m$^{-2}$. 
 
Component-level signal-to-noise is also a key factor in successful model selection.  We estimate the per-component S:N using the following proxy:
\begin{equation}
    \subtxt{S:N}_{\text{component}} = \frac{p_{\text{0, component}} \sqrt{N} }{ \sigma_{p_{0}}}
    \label{SN_proxy}
\end{equation}
where $p_{\text{0, component}}$ represents the intrinsic fractional polarization of the component, $N$ is the number of channels, and $\sigma_{p_{0}}$ represents the band-averaged noise amplitude of the data. 

In Figure~\ref{fig:BSTCproxy} we show the Component 1 signal-to-noise ratio and the $\Delta \text{RM}_{\subtxt{screen}} < 100$ rad m$^{-2}$ selection map, for the low-band population. We find that, in this regime, higher signal-to-noise is required, roughly following these 
thresholds:
\[
\begin{cases} 
30 \leq \Delta \text{RM}\leq 100 \ \text{rad m}^{-2} & \text{S:N}_{\text{component}} > 15. \\
\Delta \text{RM}\leq 30 \ \text{rad m}^{-2} & \text{S:N}_{\text{component}} > 30.
\end{cases}
\]

The results for the full-band population of two-component sources are presented in Figure~\ref{fig:combined_band_selection}d. 
The full-band improves on the low-band limits:

\[
\begin{cases} 
18 \leq \Delta \text{RM} \leq 100 \ \text{rad m}^{-2}, & \text{S:N}_{\text{component}} > 10. \\
\Delta \text{RM} \leq 18 \ \text{rad m}^{-2} & \text{S:N}_{\text{component}} > 30.
\end{cases}
\]

Through examination of the fits themselves we determine that in the small $\Delta \text{RM}_{\subtxt{screen}}$ regime where the BS model is preferred, QU-fitting fits slabs with widths as small as 5--10 rad m$^{-2}$, similar to when the BS model is fitted to the Faraday simple population. For this reason, if we conducted an experiment comparing the FS and TC models, we would observe similar results to Figure~\ref{fig:BSTCproxy}.

\subsubsection{Burn slab and internal turbulence models}
\begin{figure}
	\includegraphics[width=\columnwidth]{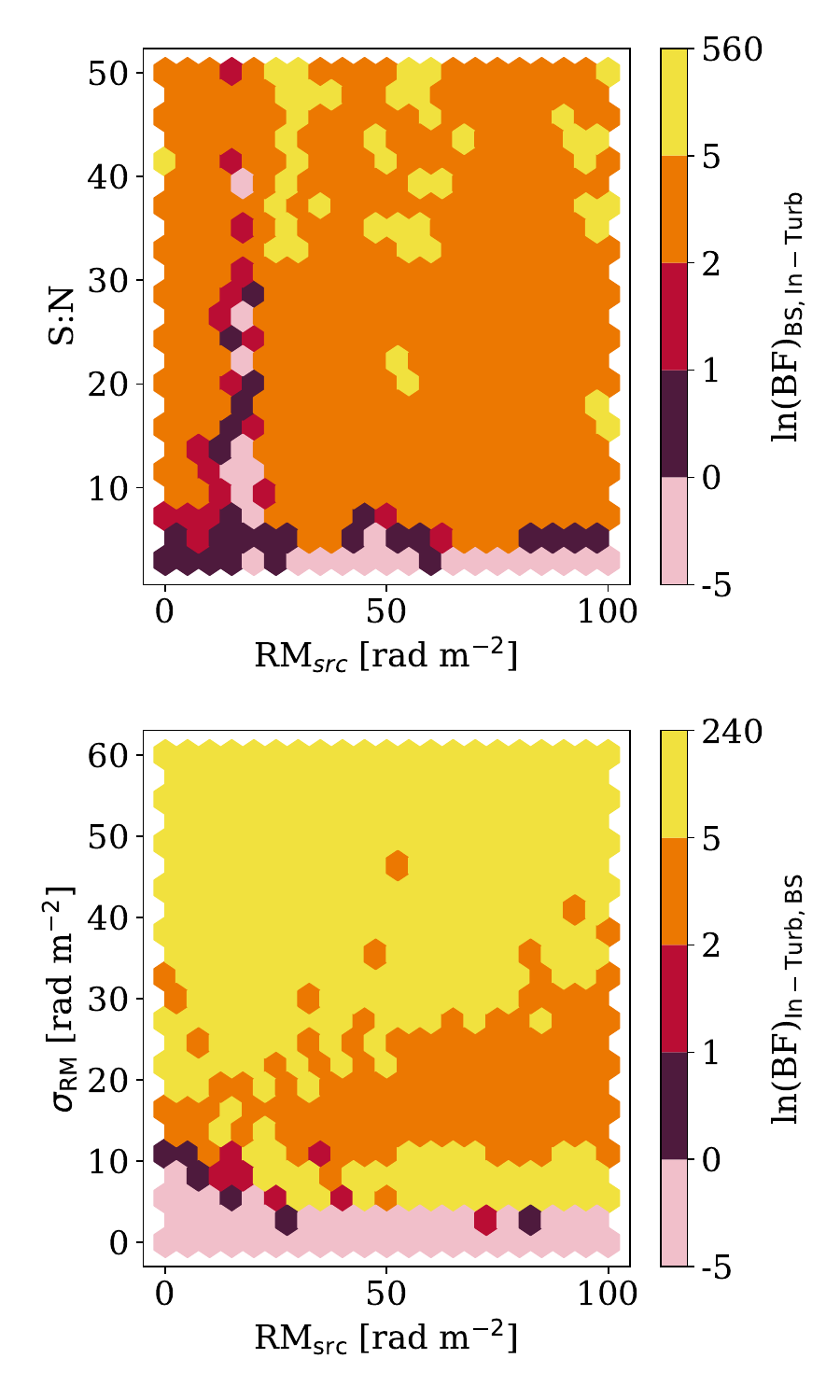}
    \caption{\textit{Top:} The parameter selection map for the POSSUM low-band (800-1088 MHz) Burn slab population when fitted with the internal turbulence model as the alternate. The map is defined by the true $\text{RM}_{\subtxt{src}}$ and S:N.  \textit{Bottom:} The parameter selection map for the low-band internal turbulence source populations when the Burn slab is the alternate model. The map is defined by the true $\text{RM}_{\subtxt{src}}$ and $\sigma_{\subtxt{RM, src}}$. For the true internal turbulence population experiment only sources with S:N > 8 are considered. Each bin is colored by the median ln(BF) value and the color bar is scaled according to Table \ref{tab:binning}.}
    \label{fig:MS_BS_band1}
\end{figure}

Our third experiment investigates the model selection performance of QU-fitting with the Burn slab (Equation \ref{BS}) and internal turbulence (Equation \ref{T-slab}) models. In Figure~\ref{fig:MS_BS_band1} we show the S:N -- $\text{RM}_{\subtxt{src}}$ parameter selection map for the low-band Burn slab population. For these models, correct model selection generally requires S:N $>$ 8.  The inconclusive classifications spike at $ \text{RM}_{\subtxt{src}} \approx 20$ rad m$^{-2}$, which corresponds to the fourth row in Figure \ref{dict_BS_MS}. At this Faraday thickness, the spectrum of the BS is similar to that of a In-Turb with $\sigma_{\subtxt{RM,src}} \approx 20$ rad m$^{-2}$ and $\text{RM}_{\subtxt{src}}$ < 15 rad m$^{-2}$. Full-band coverage partially mitigates this observational degeneracy for sources with S:N~$ >15$  (Figure~\ref{fig:combined_band_selection}c).

We determine that the most significant parameters for correct model selections for the true internal turbulence population are $\text{RM}_{\subtxt{src}}$ and $\sigma_{\subtxt{RM,src}}$, and S:N. In the bottom panel of Figure~\ref{fig:MS_BS_band1} we present the results when the BS model is the alternative. An S:N~$> 8$ is required for a reliable classification; thus, we exclude sources below this threshold from further analysis. The majority of the parameter space is recovered with strong support, with three regions of interest: 

\begin{enumerate}
\item{ The region where $\sigma_{\subtxt{RM,src}}$ < 8,}
\item{ The region 8 rad m$^{-2}$ < $\sigma_{{\mathrm{RM,src}}}$ < 18 rad m$^{-2}$ and $\mathrm{RM_{src}}$ > 25 rad m$^{-2}$}, 
\item{ The region where 25 rad m$^{-2}$ < $\sigma_{\subtxt{RM,src}}$ < 40 rad m$^{-2}$.}
\end{enumerate}

 Figure \ref{dict_BS_MS} is helpful to understand these regions. These regions are the result of these two models falling in and out of observational degeneracy with each other in the POSSUM bands. In regions (i) and (ii), the In-Turb model becomes observationally degenerate with the Burn slab, and QU-fitting tends to select the latter. Full-band data only modestly improve this limitation, and broader bandwidth is required to reliably distinguish the models.

\subsubsection{Internal turbulence and external turbulence}

\begin{figure}
	\includegraphics[width=\columnwidth]{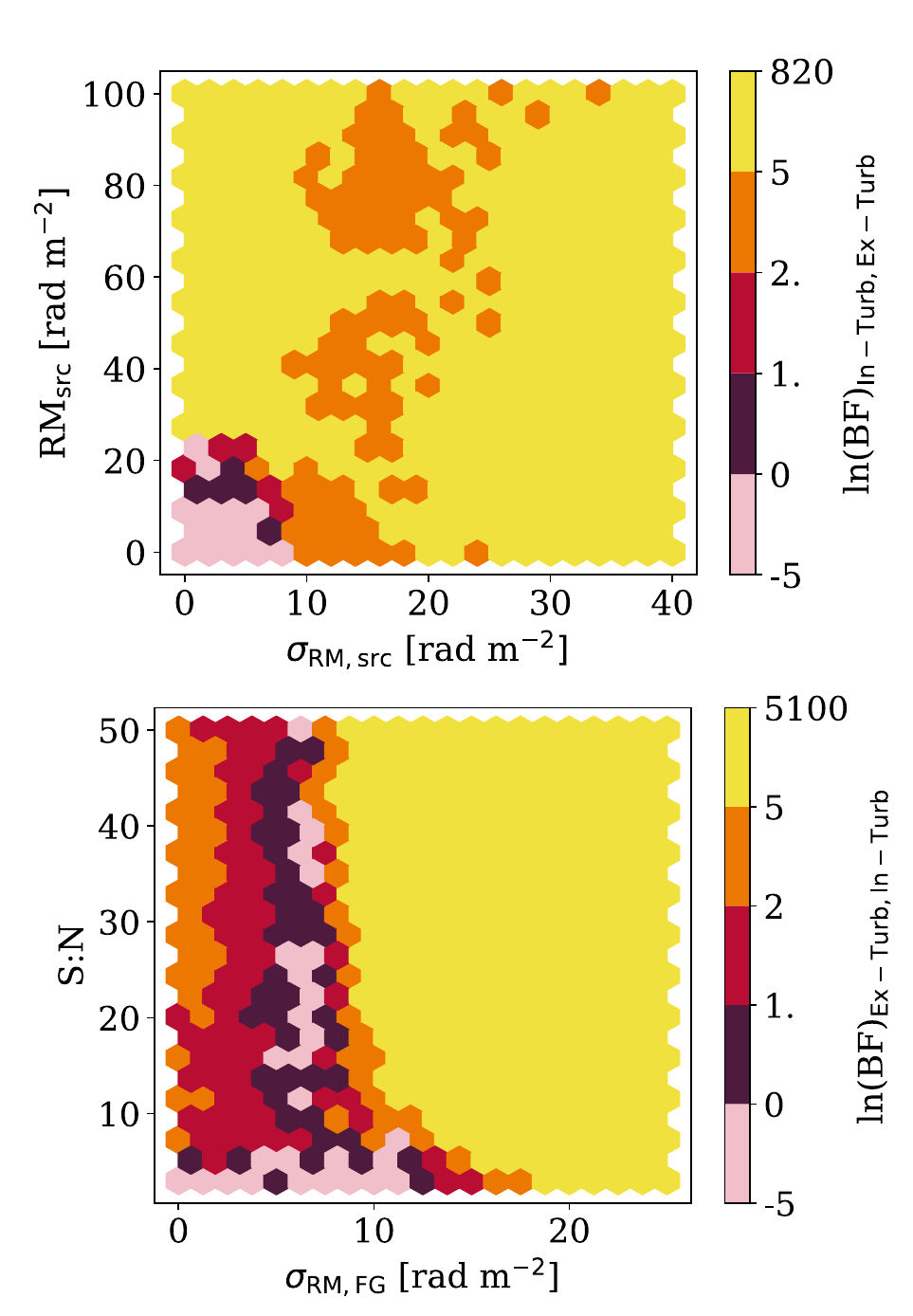}
    \caption{\textit{Top:} The parameter selection map for the true internal turbulence POSSUM low-band (800-1088 MHz) population when fit with the external turbulence model as the alternative model. This map's space is defined by the slab width ($\text{RM}_{\subtxt{src}}$) and small-scale RM dispersion parameter ($\sigma_{\subtxt{RM, src}}$). Sources  with S:N < 7 were removed from this map. \textit{Bottom:} Parameter selection map for the true external turbulence low-band population with the internal turbulence model as the alternative model. This map is defined with the S:N and small-scale foreground RM dispersion ($\sigma_{\subtxt{RM, FG}}$). Each bin is colored by the median ln(BF) value and the color bar is scaled according to Table \ref{tab:binning}}
    \label{fig:MS_RFG_src}
\end{figure}

We present the results of our experiment on the internal (Equation \ref{T-slab}, In-Turb) and external (Equation \ref{T-Screen}, Ex-Turb) turbulence models. We first consider the population of true internal turbulence sources. Proper selection for this model was found to require S:N > 7 and we consider only sources above this limit for the rest of the analysis. In the upper panel of Figure ~\ref{fig:MS_RFG_src} we show the $\sigma_{\subtxt{RM, src}}$ -- $\text{RM}_{\subtxt{src}}$ parameter map for the POSSUM low-band internal turbulence population, fit with the Ex-Turb model as an alternative. We find that QU-fitting can select the correct model with high confidence when $\text{RM}_{\subtxt{src}}$ > 20 rad m$^{-2}$ or $\sigma_{\subtxt{RM, src}}$ > 10 rad m$^{-2}$. Below these limits, the In-Turb spectra can be well approximated by an exponential, and thus QU-fitting prefers to fit a Ex-Turb model with $\sigma_{\subtxt{RM,FG}}$ < 10 rad m$^{-2}$ (as seen in the second column of Figure \ref{dict_BS_MS} and the second row of the first column of Figure \ref{dict_RFG}). 

The full-band results for the internal turbulence model are shown in Figure \ref{fig:combined_band_selection}f. We observe a slight improvement in confidence within the parameter space. However, the extra bandwidth is not enough to improve the distinguishability for source with $\text{RM}_{\subtxt{src}}$ < 20 rad m$^{-2}$ and $\sigma_{\subtxt{RM, src}}$ < 10 rad m$^{-2}$ limit, indicating that lower frequencies will be needed to select within this region. 

The second part of the experiment focuses on the true external turbulence population. We present the S:N and $\sigma_{\mathrm{RM, FG}}$ map (lower panel of Figure \ref{fig:MS_RFG_src}) to highlight the regions of parameter space where the true Ex-Turb becomes observationally degenerate with the In-Turb. Similarly to the previous experiment, we observe a spike of model indistinguishability at $5 \leq \sigma_{\subtxt{RM, FG}} \leq $ 8 rad m$^{-2}$ where QU-fitting prefers to fit the In-Turb model (with $\text{RM}_{\subtxt{src}}$ < 15 rad m$^{-2}$ and $\sigma_{\subtxt{RM, src}}$ < 20 rad m$^{-2}$). For all other parts of the parameter space, QU-fitting is able to make correct Ex-Turb selections.

The $\sigma_{\subtxt{RM, FG}} $ < 5 rad m$^{-2}$ region is still able to retain selections because the In-Turb is not directly mathematically degenerate with the Ex-Turb in this limit (see first rows of Figures~\ref{dict_BS_MS} and the first column of ~\ref{dict_RFG}). The full-band data helps break this observational degeneracy—provided that S:N > 30 — turning previously inconclusive selections into weakly confident ones.

\subsection{QU-fitting Accuracy}\label{Accuracy}

\begin{figure}
    \includegraphics[width=0.5\textwidth]{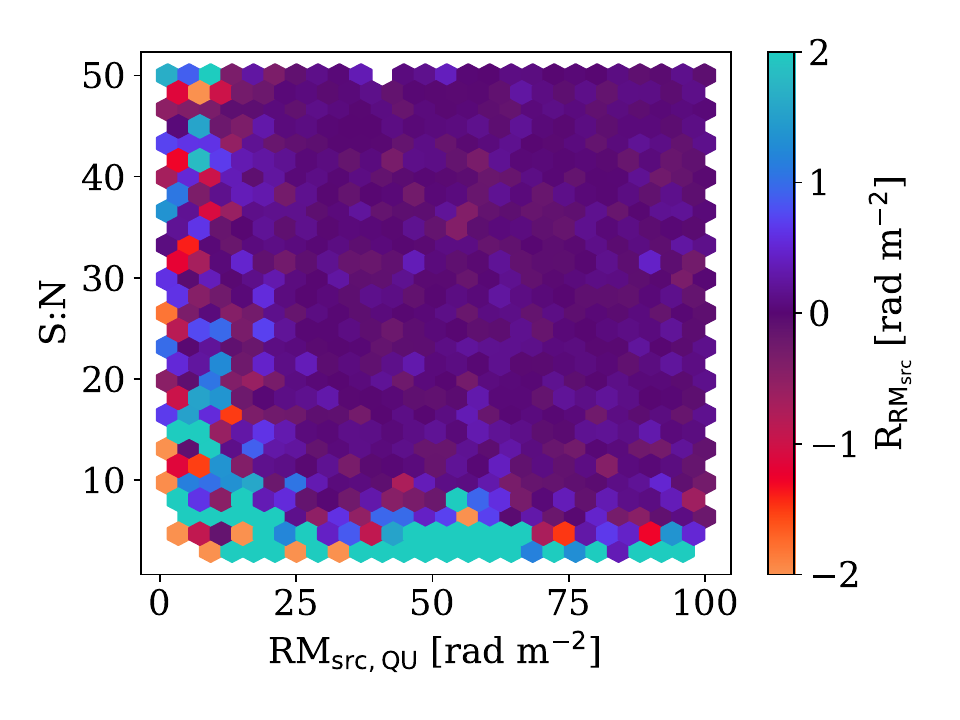}
    \caption{The parameter accuracy map for the low-band Burn slab population. The map's space is defined by the fitter's estimate for slab width $RM_{src, QU}$ and the S:N. Each bin is colored by the median residual for the population within that bin of parameter space. The color bar minimum was set to emphasize the finer trends of the plot. The true minimum of the residuals is -10 rad m$^{-2}$.}
    \label{fig:BS_error}
\end{figure}

In this section, we discuss both the accuracy of the recovered model parameters and of the measurement uncertainty produced by \texttt{RM-Tools} for the Burn slab, two-component, internal turbulence, and external turbulence models. The following results were obtained with the same source populations used in the previous section. \cite{VanEck2026} found that for the Faraday simple model, the measurement uncertainties on the model parameters accurately portray the true measurement error; we find agreement with those results (see the first row of Table \ref{all_models_error_bars}) that the measurement uncertainties for the Faraday simple model are reliable and will not discuss this model further in this work. 

In the following sections we analyze the results using the residuals defined $R = M - T$, where $T$ is the true value of the parameter and $M$ is the estimation from the QU-fitting. We denote the residual (\textit{R}) for a specific parameter with a subscript. To determine the accuracy of the measurement uncertainties produced by QU-fitting we examine the error-scaled residuals for each parameter which we define as $R$ / estimated measurement uncertainty. In the ideal case, the error-scaled residuals should have a normal distribution with zero mean and unit variance/standard-deviation. 

\begin{figure*}
    \includegraphics[width=0.9\textwidth]{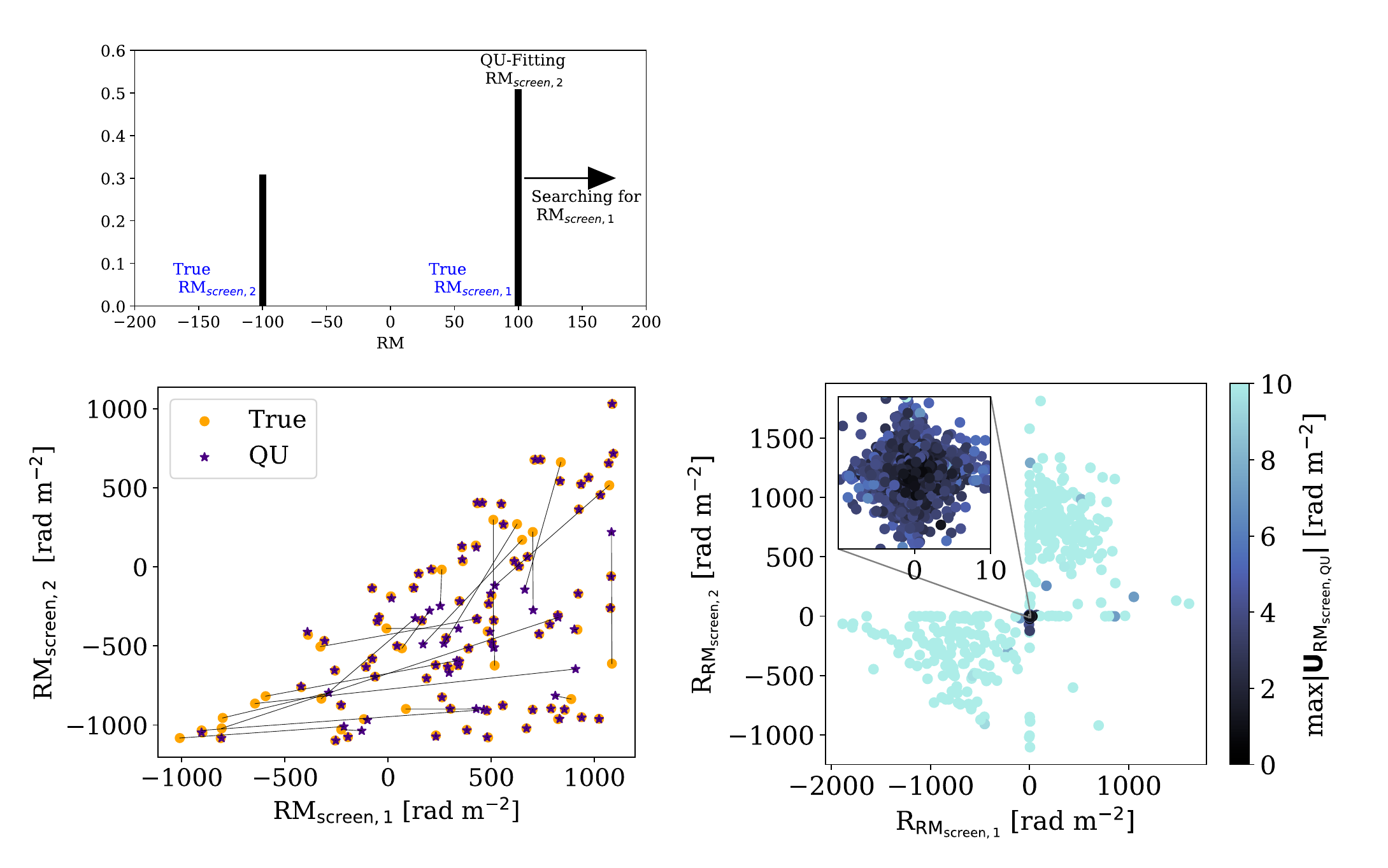}
    \caption{\textit{Left:} The map of the true (orange circles) and fitted estimates (violet stars) of the $\text{RM}_{\subtxt{screen}}$ values for random selection of the two-component low-band source population. Within the total source population there are four populations: (i) correctly fit, (ii) horizontal lines - sources where the second component is fitted correctly, (iii) vertical lines - sources where the first component is fitted correctly, (iv) diagonal lines - both components are incorrect. The panel in the upper left demonstrates how the population can come to be. QU-fitting can select the true component RM for the wrong component and then be forced to fit the other one incorrectly. \textit{Right:} Scatter plot of the RM$_{\text{screen, QU}}$ residuals of all the sources in the low-band population of the two-component model sources with both components S:N > 6. The sources are colored by the maximum absolute measurement uncertainty on $\text{RM}_{\subtxt{screen, QU}}$ between each of the two component uncertainties (max|$\mathbf{U_{{\text{RM}}_{\subtxt{screen}}}}$|). Approximately 70\% of the population is concentrated near the central region (shown in the zoom-in) of the RM residual parameter space, where both components exhibit residuals below 8 rad m$^{-2}$.}
    \label{fig:TC_error}
\end{figure*}

Similarly to the previous section, we utilize two-dimensional histograms maps, but for this part of the analysis we color the maps by median residual value within each of the bins and refer to them as parameter accuracy maps. We found that none of the goodness-of-fit metrics ($\chi^{2}$, AIC, BIC, ln(EVD)) were \textit{consistently} reliable measurements of whether or not the fitted result is an accurate estimation of the true model parameters. Instead in the following sections we explain which model parameters we found were the most impactful on the accuracy of QU-fitting.

For this analysis, we will use the QU-fitting parameter estimates instead of the true parameter values because this allows us to draw direct connections to what users will derive from real POSSUM data to the accuracy they can expect. We designate this difference by adding the subscript `QU' to the variable. We also create movement maps that show both $T$ and $M$ values on the same scatter plot. We use these maps and the spectra atlas in Appendix \ref{app:Atlas} to help us determine the cause of inaccurate parameter estimates. 

\subsubsection{Burn slab model accuracy} 

We determined the S:N and $\text{RM}_{\subtxt{src, QU}}$ values are the most important factors in establishing the accuracy of model fits. In Figure \ref{fig:BS_error} we present the parameter accuracy map for the low-band BS population defined by $\text{RM}_{\subtxt{src, QU}}$ and S:N. We observe a similar trend to that from the parameter selection maps for this model; and find that S:N > 10 and Equation \ref{FSBSrelation} provide good definition of the unbiased section of parameter space for this model. Within these bounds, the residuals are unbiased. 
We find that using the full-bandwidth improves the performance of QU-fitting, enabling unbiased recovery of BS model parameters at signal-to-noise ratios as low as S:N > 8, roughly following Equation \ref{FSBSrelationcomb}.

As a word of caution, we observe 4$\sigma$ outliers in the hexbins within the $40 < \text{RM}_{\subtxt{src, QU}} < 45$ rad m$^{-2}$ region and again at $80 < \text{RM}_{\subtxt{src, QU}} < 85$ rad m$^{-2}$, which is due to the same effect discussed in Section \ref{Models_Exp_Results}. Examples of such fits are presented in Figure \ref{fig:BF_parameter_space}b and d.

We report the mean ($\mu$) and standard deviation (\textit{std}) of the uncertainty-scaled residuals distribution for all model parameters in Table \ref{all_models_error_bars}. We conclude that for both the low-band and the full-band, the measurement uncertainties for $p_{\text{0, QU}}$, $\psi_{0, QU}$ and $\text{RM}_{\subtxt{screen, QU}}$ accurately reflect the residuals across the entire parameter space.

\subsubsection{Two-component model accuracy}\label{sec:TC_acc}
For the two-component model, we found that the determining parameters for accuracy are the true separation between the two components $\Delta \text{RM}_{\text{QU}} = \text{RM}_{\subtxt{screen, 1, QU}} - \text{RM}_{\subtxt{screen, 2, QU}}$ and the maximum measurement uncertainty between the two components (max($U_{\text{RM}_{\subtxt{screen}}}$)). We consider only sources with S:N > 6 for both components.

In the left panel of Figure \ref{fig:TC_error} we show the true parameter value and QU estimation $\text{RM}_{\subtxt{screen}}$ values for both components of 50 random sources in the low-band population. We identify four populations within our sources:
\begin{enumerate}
  \item Correctly Fitted Sources - QU-fitting has correctly estimated the RM of both components.
  \item Horizontal Lines - These source RMs are connected by horizontal lines because QU-fitting estimates the second component correctly but not the first component. 
  \item Vertical Lines - These source RMs are connected by vertical lines, because  QU-fitting correctly estimated the RM of the first component and not the second. 
  \item Diagonal Lines - These source RMs of these sources are connected by diagonal lines, meaning that QU-fitting got the RM of both components wrong.
\end{enumerate}
Population [i] consists of approximately 70 \% of the total sources, while populations [ii], [iii], and [iv] consist of the remaining 30 \%. 

These three populations come from different failure modes of the QU-fitting algorithm where only one of the true RM$_{\subtxt{screen}}$ values is captured. Populations [ii] and [iii] exist because QU-fitting finds a single component correctly, but finding the other component does not sufficiently impact the $\chi^{2}$. Population [iv] is made up of sources where QU-fitting is mislabeling. The algorithm has mistakenly selected one of the component RM peaks for the other (as demonstrated in the upper left diagram of Figure \ref{fig:TC_error}. This failure mode is caused by the rule implemented into \texttt{RM-Tools} for the TC model:
\begin{equation}
    \text{RM}_{\subtxt{screen, 1}} > \text{RM}_{\subtxt{screen, 2}}.
\end{equation}
This rule allows nested sampling to converge on a posterior. When QU-fitting incorrectly selects $\text{RM}_{\subtxt{screen, 2, QU}}$ for $\text{RM}_{\subtxt{screen, 1, QU}}$, the rule forces QU-fitting to underestimate the true $\text{RM}_{\subtxt{screen, 1}}$ and vice-versa. In these instances, QU-fitting reduces the $\chi^{2}$ sufficiently to become trapped in a local minimum and prematurely declare convergence, despite the fact that nested sampling is, in principle, designed to mitigate such failures but does not succeed here. We confirm that the residuals for $p_{\subtxt{0, QU}}$ and $\psi_{\subtxt{0, QU}}$ also follow these four trends as would be expected. 

We used a subset of 10 TC sources within each of populations [ii], [iii], and [iv] and re-fit them three times with the same QU-fitting configuration to determine how often this occurs randomly and can be corrected by rerunning. We found that, in this population, repeating the fitting provides a better result in 66 \% of cases. 

\begin{figure*}
    \includegraphics[width=0.9\textwidth]{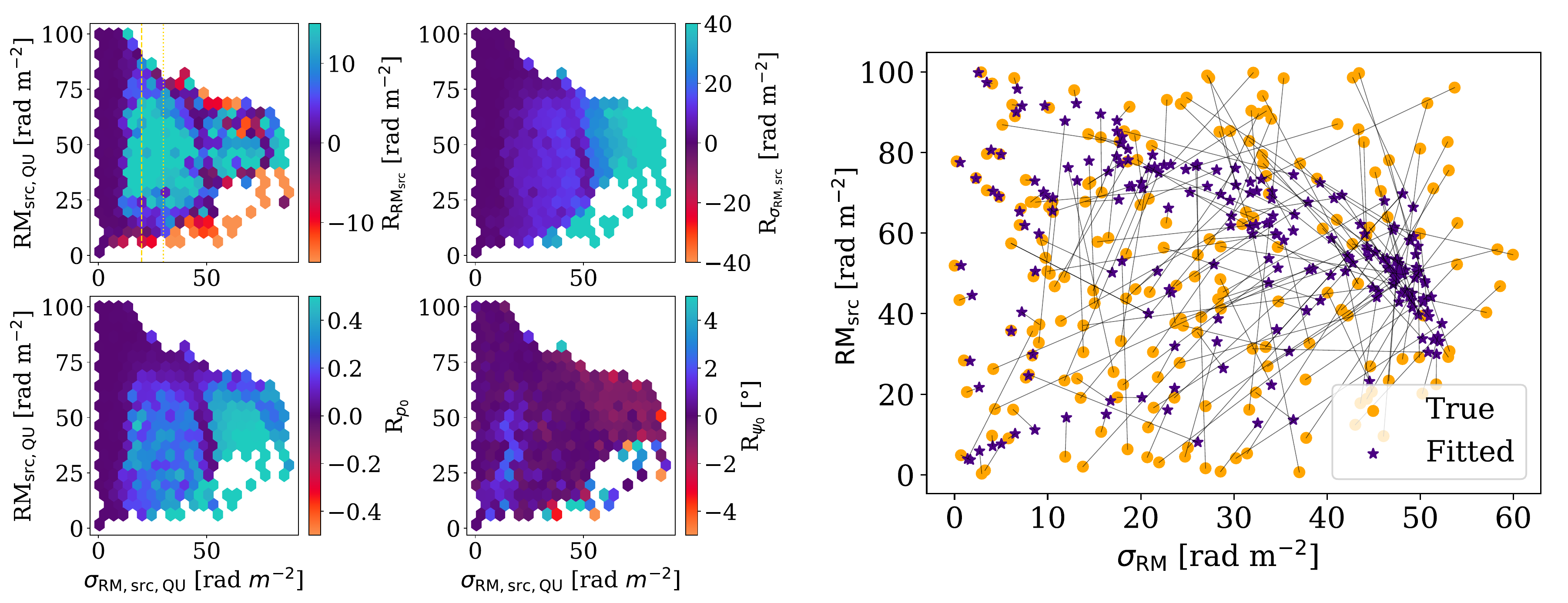}
    \caption{\textit{Left:} Four parameter accuracy maps for the internal turbulence model low-band source population (S:N > 8) We define this map with the RM$_{\subtxt{src, QU}}$ and $\sigma_{\subtxt{RM, src, QU}}$. The total population plotted here is 15 631 sources. Each bin is the median value of the residuals for $\text{RM}_{\subtxt{src, QU}}$ (top left), $\sigma_{\subtxt{RM, src, QU}} $ (top right), p$_{0}$ (bottom left), and $\psi_{0}$ (bottom right). The violet region of the maps are where QU-fitting is unbiased. The yellow dashed line in the top left map indicates the edge of this region for the POSSUM full-band. \textit{Right:} we show both the true $\sigma_{\subtxt{RM,src}}$ and $\text{RM}_{\subtxt{src}}$ value (orange circle) and the fitted values (violet star) connected by a line for a random subset of 100 sources. }
    \label{fig:MS_error}
\end{figure*}

We find that the max($U_{\text{RM}_{\text{screen, QU}}}$) between the two components is the only \textit{consistent} metric to determine if a source falls into population [i]. In the right panel of Figure \ref{fig:TC_error}, we show the residual parameter space of the RM component, with each source colored by max($U_{\text{RM}_{\text{screen, QU}}}$).  Population [i] consists of the center points (shown in the zoom in), and populations [ii] and [iii] make up the horizontal and vertical clusters, respectively. The population [iv] makes up the scattered points of the horizontal and vertical populations. We find that population [i] sources typically have:

\begin{equation}\label{TC_limits}
\text{max}( U_{\text{RM}_{\subtxt{screen}}} ) < 10 \: \text{rad m}^{-2}.
\end{equation}

In this regime, parameter residuals have biases ranging from: $\text{RM}_{\subtxt{screen, QU}}$: +5 rad m$^{-2}$, $p_{0, \text{QU}}$: +0.02, and $\psi_{0, \text{QU}}$: $\pm$ 22$^{\circ}$. Using the full-band does not significantly impact the amplitude of the residuals but slightly decreases the component S:N level required to obtain this result and increases the percentage of the sources in population [i] to 74\%. 

The uncertainty measurements are not a reliable approximation of the true residual for any of the parameters and frequency regimes. When we only consider population [i] sources (shown in Table \ref{all_models_error_bars}), the mean and std — while significantly decreased for $p_{0, \text{QU}}$, $\psi_{0, \text{QU}}$, and $\text{RM}_{\subtxt{screen, QU}}$ compared to the full population, are still not reflective of the true residual distributions. So, while small values of max( $U_{\text{RM}_{\subtxt{screen}}})$ can help pinpoint unbiased results, they do not reflect the true uncertainties.

\subsubsection{Internal turbulence model accuracy}
The significant parameters that determine the accuracy of In-Turb model fits are S:N,  $\text{RM}_{\subtxt{src, QU}}$ and $\sigma_{\subtxt{RM, src, QU}} $. Due to the larger number of influential parameters for this model's accuracy, we started with a source population of 18000. We found that sources with S:N < 8 have parameter estimates that deviate from their true values by approximately 3$\sigma$, and were removed from the rest of this analysis. We are then left with 15 631 sources in the low-band POSSUM population and 15 673 sources in the full-band population.

In the left panel of Figure \ref{fig:MS_error}, we show the structure of the residuals $\text{RM}_{\subtxt{src, QU}}$, $\sigma_{\subtxt{RM, src, QU}} $, $p_{0, \text{QU}}
$, and $\psi_{0, \subtxt{QU}}$ for the low-band population. We conclude that for low-band, only sources with $\sigma_{\subtxt{RM, src, QU}}$ < 18 rad m$^{-2}$ are accurately fit with this model. In this region we find that the residual distributions are unbiased. For the full-band we find that QU-fitting can provide reasonable parameter estimates for S:N > 8 and  $\sigma_{\subtxt{RM, src, QU}}$ < 25 rad m$^{-2}$ (this is also shown in the yellow line). In this region, the residual biases remain the same as for the low-band.

In the regime beyond the $\sigma_{\subtxt{RM, src, QU}}$ > 18 rad m$^{-2}$ limit, the residuals become much larger and the structure more complex. We find that QU-fitting prefers to overestimate $\sigma_{\subtxt{RM, src, QU}}$ and fit $\text{RM}_{\subtxt{src, QU}} \approx 50$ rad m$^{-2}$ (as shown in the right panel of Figure \ref{fig:MS_error}). This behavior of fitting $\text{RM}_{\subtxt{src, QU}}$ `to the middle' is triggered by the true $\sigma_{\subtxt{RM, src}}$ crossing a threshold beyond which, for POSSUM frequencies, it becomes observationally degenerate with $\text{RM}_{\subtxt{src}}$. This is supported by the QU spectra in the 5-10 rows and the second column of Figure \ref{dict_BS_MS}. There is a specific `bump' feature in the first column of spectra that helps QU-fitting lock onto an accurate fit. For our POSSUM low-band population that feature is no longer in the spectra when $\sigma_{\subtxt{RM, src,QU}} \approx 15$ rad m$^{-2}$. We conclude that after this threshold the spectra at the high and low extremes of the $\text{RM}_{\subtxt{src}}$ range are degenerate with the  mid-range ($\text{RM}_{\subtxt{src}} \approx 50$ rad m$^{-2}$) and that QU-fitting will fit them all with the same $\text{RM}_{\subtxt{src}}$.

In this regime this model experiences significant degeneracy between $\text{RM}_{\subtxt{src, QU}}$ and $\sigma_{\text{RM, src, QU}}$ and QU-fitting can produce large residuals. $\text{RM}_{\subtxt{src, QU}}$ and $\sigma_{\subtxt{RM, src, QU}}$ both deviate by up to $\pm$ 30 rad m$^{-2}$. The other model parameters also have large residuals. From the results presented in Table \ref{all_models_error_bars}, we confirm that in both regimes QU-fitting overestimates $p_{0}$, $\text{RM}_{\subtxt{src}}$, and $\sigma_{\subtxt{RM, src}}$. The magnitude of the residuals for these parameters decreases for the full-band because QU-fitting is able to break the observational degeneracy for larger $\sigma_{\subtxt{RM, src}}$.

For the POSSUM low-band regime, we conclude that none of the parameters have uncertainty estimates calculated by QU-fitting that are representative of the true residuals. For full-band sources, QU-fitting is able to provide uncertainty measurements for $\sigma_{\subtxt{RM, src}}$ that coincide with the true residuals with little variance. These results consider the full model parameter space.

\subsubsection{External turbulence model accuracy}\label{T-Screenacc}

\begin{figure*}
    \includegraphics[width=0.9\textwidth]{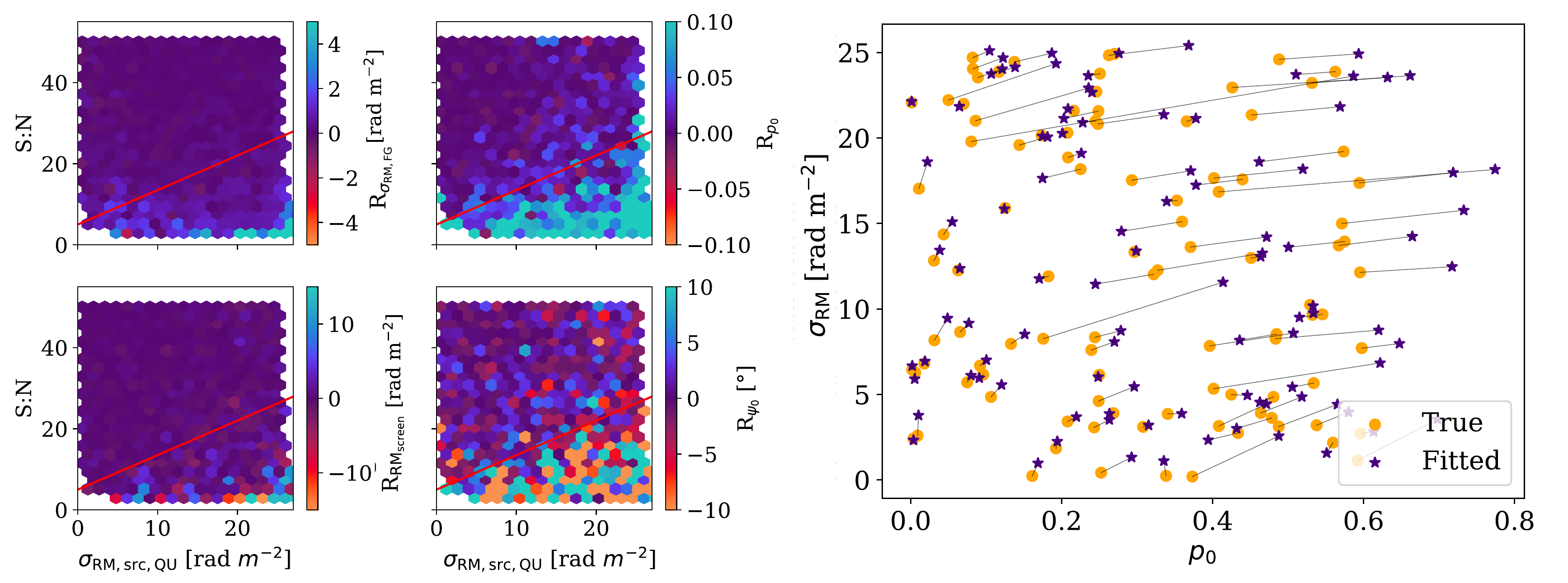}
    \caption{\textit{Left:} Four parameter accuracy maps for the external turbulence model low-band population. These maps are defined by QU-fitting's estimation for the foreground turbulence $\sigma_{\subtxt{RM, FG, QU}} $ and S:N parameter maps. Each bin is colored by the median residuals for all four model parameters: $\sigma_{\subtxt{RM, FG}} $ (upper left), $p_{0}$ (upper right), $\text{RM}_{\subtxt{screen}}$ (lower left), and $\psi_{0}$ (lower right). \textit{Right:} We show the location of the true $p_{0}$ and $\sigma_{\subtxt{RM, FG}}$ values (orange circle) and the fitted values (violet star) for a random subset of 100 sources. It is important to note that QU-fitting is increasing the fractional polarization to create the best fit.}
    \label{fig:RFG_error}
\end{figure*}

The primary factors determining accuracy for the Ex-Turb model are the S:N and $\sigma_{\subtxt{RM, FG, QU}} $ values. From the structure of the parameters' residuals of the low-band regime (left panel of Figure \ref{fig:RFG_error}), we define the region of parameter space with inequalities:
\begin{equation}\label{RFG_accuracy_relation}
    \subtxt{S:N} > 0.65 \sigma_{\subtxt{RM, FG, QU}}  + 4
\end{equation}
to be reasonably unbiased. Using the full POSSUM band data improves the accuracy of QU-fitting with this model. We observe that with the full-band, the S:N limit becomes:
\begin{equation}
    \subtxt{S:N} > 0.40 \sigma_{\subtxt{RM, FG, QU}}  + 4.
    \label{T-Screen_accuracy_comb}
\end{equation}
Below these S:N boundaries, QU-fitting overestimates $\sigma_{\subtxt{RM, FG, QU}}$, $p_{0, \subtxt{QU}}$, and $\text{RM}_{\subtxt{screen}}$ by $\pm$ 10 rad m$^{-2}$. From these results and visual inspection of the posterior distributions, we attribute this effect to the QU-fitting process preferring to misestimate $p_{0, QU}$ rather than change its estimate for $\sigma_{\mathrm{RM, FG, QU}}$ (see the right panel of Figure \ref{fig:RFG_error}). We theorize that because this model depolarizes quickly, QU-fitting is more inclined to get reasonably close to the correct $\sigma_{\mathrm{RM, FG}}$ and then increase the fractional polarization "to stretch the exponential" and improve the $\chi^{2}$. 

To determine how to help QU-fitting overcome this  behavior, we performed an additional set of experiments with a sample of 5000 Ex-Turb sources with S:N > 10, $\text{RM}_{\subtxt{screen}}$: 0-100 rad m$^{-2}$, and $\sigma_{\subtxt{RM,FG}}$: 0-25 rad m$^{-2}$. We QU-fit the sources using the Ex-Turb model in three experiments: (i) a null case, implemented identically to the original fitting procedure; (ii) a case in which the prior range for $\text{RM}_{\subtxt{screen}}$ was restricted to 0–150 rad m$^{-2}$; and (iii) a case in which the number of sampling points used to explore the parameter space was increased by a factor of 1.5. We choose these tests because they are changes that a user would reasonably be able to do with the samplers and QU-fitting configuration. We found that both changes marginally improved the performance of the QU-fitting. While we suggest that the user implement these changes for this model, they should expect these issues to persist in their results. We note that both of these behaviors are more significantly reduced by using the full-band. This is likely because the added bandwidth helps to constrain $\sigma_{\mathrm{RM, FG, QU}}$.

\begin{table*}
\centering
\caption{Mean ($\mu$) and standard deviation (std) of the error-scaled residual distributions for four of the models, for both POSSUM low-band and full-band populations. Parameters that have a standard deviation close to unity means that QU-fitting produces measurement uncertainties that closely reflect the true error. The error on the std was calculated as $\text{SE} = \frac{std}{\sqrt{2(N-1)}}$ where N is the number of sources. The values for two-component model only consist of Population [i] sources as defined in Section \ref{sec:TC_acc} to be sources where QU-fitting avoids the failure modes for this model.}
\label{all_models_error_bars}
\begin{tabular}{|l|cc|cc|}
\hline

\multicolumn{5}{|c|}{\textbf{Faraday simple Model}} \\ \hline
Parameter & Low-band $\mu$ & Low-band std & Full-band $\mu$ & Full-band std \\ \hline
$p_{0}$ & $-0.002 \pm 0.001$ & $1.0 \pm 0.01$ & $-0.003 \pm 0.001$ & $1.0 \pm 0.01$ \\
$\psi_{0}$ [$^\circ$] & $0.05 \pm 0.01$ & $1.06 \pm 0.01$ & $0.03 \pm 0.01$ & $1.05 \pm 0.01$ \\
$\text{RM}_{\subtxt{screen}}$ [rad m$^{-2}$] & $-0.03 \pm 0.01$ & $1.1 \pm 0.01$ & $- 0.02 \pm 0.01$ & $1.1 \pm 0.01$ \\ \hline

\multicolumn{5}{|c|}{\textbf{Burn Slab Model}} \\ \hline
Parameter & Low-band $\mu$ & Low-band std & Full-band $\mu$ & Full-band std \\ \hline
$\text{RM}_{\subtxt{src}}$ [rad m$^{-2}$] & $0.13 \pm 0.02$ & $2.2 \pm 0.02$ & $0.29 \pm 0.01$ & $1.1 \pm 0.01$ \\
$p_{0}$ & $-0.003 \pm 0.008$ & $1.1 \pm 0.01$ & $-0.003 \pm 0.008$ & $1.0 \pm 0.01$ \\
$\psi_{0}$ [$^\circ$] & $0.003 \pm 0.01$ & $1.2 \pm 0.001$ & $0.03 \pm 0.01$ & $1.1 \pm 0.01$ \\
$\text{RM}_{\subtxt{screen}}$ [rad m$^{-2}$] & $-0.03 \pm 0.01$ & $1.1 \pm 0.01$ & $0.02 \pm 0.01$ & $1.1 \pm 0.01$ \\ \hline

\multicolumn{5}{|c|}{\textbf{Two-Component Model}} \\ \hline
Parameter & Low-band $\mu$ & Low-band std & Full-band $\mu$ & Full-band std \\ \hline
$p_{0}$ & $0.20 \pm 0.02$ & $1.94 \pm 0.01$ & $0.38 \pm 0.05$ & $1.80 \pm 0.01$ \\
$\psi_{0}$ [$^\circ$] & $-0.008 \pm 0.6$ & $2.75 \pm 0.04$ & $-0.003 \pm 0.09$ & $2.58 \pm 0.06$ \\
$\text{RM}_{\subtxt{screen}}$ [rad m$^{-2}$] & $0.39 \pm 0.8$ & $8.5 \pm 0.6$ & $0.3 \pm 0.9$ & $8.1 \pm 0.6$ \\ \hline

\multicolumn{5}{|c|}{\textbf{Internal Turbulence Model}} \\ \hline
Parameter & Low-band $\mu$ & Low-band std & Full-band $\mu$ & Full-band std \\ \hline
$p_{0}$ & $0.48 \pm 0.01$ & $1.21 \pm 0.007$ & $0.45 \pm 0.01$ & $1.46 \pm 0.01$ \\
$\psi_{0}$ [$^\circ$] & $-0.13 \pm 0.01$ & $1.75 \pm 0.01$ & $-0.11 \pm 0.01$ & $1.70 \pm 0.01$ \\
$\text{RM}_{\subtxt{screen}}$ [rad m$^{-2}$] & $-0.11 \pm 0.03$ & $3.82 \pm 0.02$ & $-0.10 \pm 0.03$ & $3.14 \pm 0.02$ \\
$\text{RM}_{\subtxt{src}}$ [rad m$^{-2}$] & $0.27 \pm 0.02$ & $1.82 \pm 0.01$ & $0.15 \pm 0.01$ & $1.22 \pm 0.08$ \\
$\sigma_{\text{RM, src}}$ [rad m$^{-2}$] & $0.36 \pm 0.02$ & $2.52 \pm 0.01$ & $0.31 \pm 0.01$ & $1.11 \pm 0.008$ \\ \hline

\multicolumn{5}{|c|}{\textbf{External Turbulence Model}} \\ \hline
Parameter & Low-band $\mu$ & Low-band std & Full-band $\mu$ & Full-band std \\ \hline
$p_{0}$ & $0.55 \pm 0.01$ & $0.98 \pm 0.009$ & $0.18 \pm 0.01$ & $0.92 \pm 0.01$ \\
$\psi_{0}$ [$^\circ$] & $0.02 \pm 0.1$ & $1.4 \pm 0.01$ & $0.02 \pm 0.01$ & $1.3 \pm 0.01$ \\
$\text{RM}_{\subtxt{screen}}$ [rad m$^{-2}$] & $0.11 \pm 0.02$ & $1.94 \pm 0.01$ & $-0.002 \pm 0.01$ & $1.02 \pm 0.01$ \\
$\sigma_{\text{RM, FG}}$ [rad m$^{-2}$] & $0.41 \pm 0.3$ & $4.3 \pm 0.5$ & $0.38 \pm 0.01$ & $1.19 \pm 0.01$ \\ \hline
\end{tabular}
\end{table*}

We report the mean and standard deviation of the uncertainty-scaled residual distributions for both bandwidth populations in Table~\ref{all_models_error_bars}. For the low-band only the uncertainties on the fractional polarization ($p_{0,\mathrm{QU}}$) reliably reflect the true residuals, with standard deviations close to unity. The full-band results show that the measurement uncertainties for $p_{0,\mathrm{QU}}$ and $\text{RM}_{\subtxt{screen, QU}}$  are consistent with the observed residuals.

\section{Discussion of Results}\label{Discussion}
In the following subsections we discuss some of our experiment's results in the context of applying QU-fitting to real data. We first present a table of guidelines for using QU-fitting on the models in this work. Secondly, we elaborate on our results for the two-component model by taking the previous literature into context. Third, we will interpret our results for the internal and external turbulence models. Finally, we broaden our guide to include predictions for other surveys.

\begin{table*}[H]
\centering
\caption{Summary of guidelines for QU-fitting using both POSSUM low-band and full-band data. We provide separate guidelines for model selection and model accuracy as they can be different.  The boundaries for the model accuracy (Section \ref{Accuracy}) define the regions of parameter space where the accuracy of the model estimation from QU-fitting is unbiased. The accuracy guidelines apply to all model parameters except for $\psi_{\subtxt{0, QU}}$, which experiences biases and inaccuracies that are not reliably predictable.}

\label{Conclusion_Table}
\begin{tabular}{|l|l|l|}
\hline
\textbf{Model} & \textbf{POSSUM low-band} & \textbf{POSSUM full-band} \\
\hline
\multicolumn{3}{|c|}{\textbf{Model Selection}} \\
\hline
Faraday simple (FS) & \begin{tabular}[t]{@{}l@{}} S:N $>$ 7\end{tabular} & \begin{tabular}[t]{@{}l@{}}S:N $>$ 7 \end{tabular}\\
\hline
Burn slab (BS) & \begin{tabular}[t]{@{}l@{}}
S:N $>$ 7\\ 
AND \\
S:N $>$ $-2.8\, \text{RM}_{\subtxt{src}} + 70$\end{tabular} &
     \begin{tabular}[t]{@{}l@{}}S:N $>$ 7  \\ 
     AND \\
     S:N $>$ $-3.8\, \text{RM}_{\subtxt{src}} + 70$\end{tabular} \\
\hline
Two-component (TC) & \begin{tabular}[t]{@{}l@{}}
     if $\Delta$\text{RM}$>$ 100 rad m$^{-2}$ then S:N$_{\text{component}}$ $>$ 6 \\
     if 100 $\geq$ $\Delta$ \text{RM}$\geq$ 30 rad m$^{-2}$ then S:N$_{\text{component}}$ $>$ 15 \\
     if $\Delta$\text{RM}$\leq$ 30 rad m$^{-2}$ then S:N$_{\text{component}}$ $>$ 30 
     \end{tabular} &
     \begin{tabular}[t]{@{}l@{}}
     if $\Delta$\text{RM}$>$ 100 rad m$^{-2}$ then S:N$_{\text{component}}$ $>$ 7 \\
     if 100 $\geq$ $\Delta$\text{RM}$\geq$ 18 rad m$^{-2}$ then S:N$_{\text{component}}$ $>$ 10  \\
     if $\Delta$\text{RM}$\leq$ 18 rad m$^{-2}$ then S:N$_{\text{component}}$ $>$ 30
     \end{tabular} \\
\hline
Internal Turbulence (In-Turb) & \begin{tabular}[t]{@{}l@{}}
    S:N $>$ 8 \\
    AND \\
    if $\text{RM}_{\subtxt{src}} > 10$ then $\sigma_{\text{RM},\text{src}} > 8$ rad m$^{-2}$ \\
    if $\text{RM}_{\subtxt{src}} < 10$ then $\sigma_{\text{RM},\text{src}} > 20$ rad m$^{-2}$
    \end{tabular} &
    \begin{tabular}[t]{@{}l@{}}
    S:N $>$ 8 \\
    AND \\
    if $\text{RM}_{\subtxt{src}} > 10$ then $\sigma_{\text{RM},\text{src}} > 7$ rad m$^{-2}$ \\
    if $\text{RM}_{\subtxt{src}} < 10$ then $\sigma_{\text{RM},\text{src}} > 20$ rad m$^{-2}$
    \end{tabular} \\
\hline
External Turbulence (Ex-Turb) & \begin{tabular}[t]{@{}l@{}}
    S:N $>$ 7 \\
    AND \\
    $\sigma_{\text{RM},\text{FG}} < 8$ or $\sigma_{\text{RM},\text{FG}} > 12$ rad m$^{-2}$
    \end{tabular} &
    \begin{tabular}[t]{@{}l@{}}
    S:N $>$ 7 \\
    OR \\
    if $\sigma_{\text{RM},\text{FG}} > 8$ rad m$^{-2}$ then S:N $>$ 25
    \end{tabular} \\
\hline
\multicolumn{3}{|c|}{\textbf{Model Accuracy}} \\
\hline
FS & \begin{tabular}[t]{@{}l@{}} Not investigated see Section \ref{Accuracy}. \end{tabular} & \begin{tabular}[t]{@{}l@{}} Not investigated, see Section \ref{Accuracy}. \end{tabular}\\
\hline
BS & \begin{tabular}[t]{@{}l@{}}
     S:N $>$ 8\\
     AND \\
     $\text{RM}_{\subtxt{src}} > 18$ rad m$^{-2}$
     \end{tabular} &
     \begin{tabular}[t]{@{}l@{}}
     S:N $>$ 7\\
     AND \\
     $\text{RM}_{\subtxt{src}} > 15$ rad m$^{-2}$
     \end{tabular} \\
\hline
TC & \begin{tabular}[t]{@{}l@{}}
     S:N$_{\text{component}}$ $>$ 6 \\
     AND \\
     $\max(\mathbf{U}_{\text{RM}_{\subtxt{screen}}}) < 10$ rad m$^{-2}$
     \end{tabular} &
     \begin{tabular}[t]{@{}l@{}}
     S:N$_{\text{component}}$ $>$ 6 \\
     AND \\
     $\max(\mathbf{U}_{\text{RM}_{\subtxt{screen}}}) < 10$ rad m$^{-2}$
     \end{tabular} \\
\hline
In-Turb & \begin{tabular}[t]{@{}l@{}} $\sigma_{\text{RM, src, QU}}$ < 20  rad m$^{-2}$ \end{tabular} & \begin{tabular}[t]{@{}l@{}} $\sigma_{\text{RM, src, QU}}$  < 30 rad m$^{-2}$ \end{tabular} \\
\hline
Ex-Turb & \begin{tabular}[t]{@{}l@{}} for $p_0$, $\text{RM}_{\subtxt{screen}}$, $\sigma_{\text{RM,FG}}$ follow Equation~\ref{RFG_accuracy_relation} \end{tabular} & \begin{tabular}[t]{@{}l@{}}
$p_0$, $\text{RM}_{\subtxt{screen}}$, $\sigma_{\text{RM,FG}}$ follow Equation~\ref{T-Screen_accuracy_comb}  \end{tabular} \\
\hline
\end{tabular}
\end{table*}

\subsection{Overview of QU-fitting General Guidelines}
We present our set of guidelines to advise model selection (upper section) and accuracy of the fits (bottom section) for the POSSUM data in Table \ref{Conclusion_Table}. These guidelines  are derived from the different experiments we conducted. For each model, the details on their derivations are as follows:
\begin{itemize}
    \item Faraday simple - This model is dominated by a S:N dependence, QU-fitting can confidently select this model within the S:N defined by the BS:FS experiment. We expect this to be true for all other models as well.  
    \item Burn slab - This result is the combination of the BS:FS, BS:TC, and BS:In-Turb experiments. The spike observed of inconclusive selections in the BS:In-Turb and the inconclusive region in the BS:TC experiment are within the inconclusive selection zone defined by the BS:FS limits (Equation \ref{FSBSrelation} and Equation\ref{FSBSrelationcomb}). QU-fitting may elevate the goodness-of-fit metrics for the FS and TC when the nodes of the slab's sinc functions align with the edges of the frequency range (ie. Figure  \ref{fig:BF_parameter_space}). This may become more important in the case of significant missing channels and is not reflected in these guidelines.  
    \item Two-component - These guidelines take the FS:BS and TC:BS experiments into account.  We estimate that a higher  S:N$_{\subtxt{component}}$ limit than that derived from the TC:BS experiment is necessary in the majority of the parameter space to account for selection with the FS model. In the small component separation limit, the threshold derived in Section \ref{BS_TC} is suitable. 
    \item In-Turb – The two experiments (BS:In-Turb and In-Turb:Ex-Turb) yield overlapping exclusion zones. This guideline is the result of combining them. Since QU-fitting frequently fits Ex-Turbmodels with small $\sigma_{\mathrm{RM, FG}}$, we infer that the exclusion zones derived from the In-Turb:Ex-Turb experiment is a good approximation for FS and TC model fitting as well.
    \item Ex-Turb - These guidelines reflect both the results from the Ex-Turb:In-Turb experiment and also the Ex-Turb:FS mathematical degeneracy in small $\sigma_{\subtxt{RM, FG}} $ regime (utilizing the results from the FS:BS experiment). 
\end{itemize}

The recommendations for obtaining an accurate model are taken directly from the results (Section \ref{Accuracy}) with little to no change. 

\subsection{Discussion of Two Component Model Results}
Our results for the TC model are closely aligned with the previous work by \cite{Miyashita2019}. While not directly comparable, they are consistent in that under the limit $\Delta \text{RM}_{\subtxt{screen}}$ < 100 rad m$^{-2}$ we find that the S:N required for confident selection increases, as QU-fitting prefers to fit thin-slab ($\text{RM}_{\subtxt{src, QU}}$ < 15 rad m$^{-2}$) to
the spectra. We suspect this result indicates that QU-fitting has the potential for precise component fitting in the S:N < 8 range.

The relationship between our results and the POSSUM rotation measure spread function (RMSF) is also worth considering. The low-band and full-band RMSFs are $58~\mathrm{rad\ m^{-2}}$ and $31~\mathrm{rad\ m^{-2}}$, respectively \citep{Gaensler2025}. Although resolution influences model discrimination, we do not find a straightforward linear relationship between RMSF narrowing and TC selectability. For both band populations, in the instances where QU-fitting favors the BS model over the TC model, the fits often correspond to slabs with $\text{RM}_{\subtxt{src, QU}}$ thicknesses of less than $20~\mathrm{rad\ m^{-2}}$. In addition, component signal-to-noise also remains a critical factor in determining model preference.

In evaluating the accuracy of QU-fitting with the two-component model, we identify three distinct source populations that arise from two failure modes of the \texttt{RM-Tools} QU-fitting algorithm. We show that the maximum uncertainty in rotation measure between fitted components provides a robust and consistent diagnostic of these failure modes. We therefore recommend, when using this model, performing multiple independent QU-fitting realizations; solutions exhibiting large component RM uncertainties should be interpreted as having converged to a failure mode and discarded or refit.

\subsection{Discussion of Ex-Turb and In-Turb Model Results}

The challenge of distinguishing between the internal and external turbulence models is well documented by \cite{OSullivan2012}, \cite{Anderson2018}, and \cite{Kaczmarek2018}. Our results for model selection between these two models using Bayesian evidence and POSSUM data are encouraging. Our analysis suggests that both the POSSUM low- and full-band data can reliably differentiate between internal and external turbulence in a majority of the parameter space, representing a notable step forward distinguishing these two forms of complexity.

Recent work by \cite{Stil2025} describes the systematic quantifiable discrepancy between the perfect turbulent models (specifically Equation \ref{T-Screen}) and what a real source surrounded by a turbulent medium and conformed to a beam could be. Their results are complementary to this work in that here we assume that the source is a true external or internal turbulent model as defined by Equation \ref{T-Screen} and \ref{T-slab} respectively. Understanding how QU-fitting will handle non-Gaussian models is beyond the scope of this work. However, users should also be aware that the effects described by \cite{Stil2025} will be present in their results.

Accomplishing accurate model estimates for the internal turbulence model for the majority of the model parameter space remains out of reach of POSSUM low- and full-band data. Due to observational degeneracy that this model has with itself in these bands, the fitting yields an accurate recovery of parameters only within a limited region of the parameter space (as defined in Table \ref{Conclusion_Table})

\subsection{Implications for Other Surveys}
The results we have are directly compatible only with the POSSUM survey specifications. In this section we discuss the QU-fitting potential of other similar surveys. We summarize these survey frequency bands and channel configurations in Table \ref{tab:survey_specs}.

The Spectral and Polarisation in Cutouts of Extragalactic sources from Rapid ASKAP Continuum Survey (SPICE-RACS) in its entirety will have access higher frequency data than POSSUM but with coarser channel spacings. We predict that the low-band and mid-band together will achieve similar results to Table \ref{Conclusion_Table}. When using the entire band (including RACS-high), this survey will be able to achieve fits for the internal turbulence model with accuracy surpassing the POSSUM $\sigma_{\text{RM, src}}$ = 35 rad m$^{-2}$ limit. We anticipate that the full-band may be able to improve the TC model's selectability in the low $\Delta \text{RM}_{\subtxt{screen}}$ regime. For the Ex-Turb model we predict that this extra bandwidth will help QU-fitting overcome the degeneracy with models like the Burn slab and the internal turbulence model, but are not able to predict if it will help constrain the $p_{0}-\sigma_{\subtxt{RM, FG}}$ behavior of the fractional polarization seen in the external turbulence model (see Section \ref{T-Screenacc}).

A possible full-spectral-resolution polarized processing of the VLASS survey would have the channel resolution and higher frequency regimes necessary to surpass the $\sigma_{\subtxt{RM, src, QU}} $ = 30 rad m$^{-2}$ POSSUM accuracy limit for the internal turbulence model. We also predict that it will be able to better constrain the $p_{0}-\sigma_{\subtxt{RM, FG}}$ effect in the external turbulence model and make improvements to Burn slab selectability in the thin slab regime.

We anticipate that the MIGHTEE survey and The HI/OH/Recombination line survey of the Milky Way (THOR) with their higher frequency range and fine channel spacings (see Table \ref{tab:survey_specs}), will perform similarly to the full POSSUM survey (with RACS-high). With the exception that it will be able to observe Ex-Turb sources with larger $\sigma_{\subtxt{RM, FG}}$ than POSSUM but should still encounter the model's observational degeneracy. 

\begin{table}
\centering
\begin{tabular}{|p{2.8cm}|p{4cm}|}
\hline
\multicolumn{2}{|c|}{\textbf{Other survey specifications}} \\
\hline
\textbf{Survey Name} & \textbf{Coverage / Specifications} \\
\hline
VLASS & 2--4 GHz with 128 channels \citep{Mao2014} \\
THOR & 1--2 GHz with 512 channels \citep{Beuther2016} \\
SPICE-RACS & low-band: 744-1032 MHz, mid-band: 1295.5-1439.5 MHz, high-band: 1511.1-1799.5 MHz  with 1 MHz spacings \citep{Duchesne2025, Duchesne2023, Hale2021, McConnell2020} \\
MIGHTEE & 856--1712 MHz with 0.026 MHz channel widths \citep{Jarvis2016} \\
\hline
\end{tabular}
\caption{Specifications of other surveys that may yield similar or potentially better results for QU-fitting than those discussed here. Included are the MeerKAT International GHz Tiered Extragalactic Exploration (MIGHTEE),  Spectral and Polarisation in Cutouts of Extragalactic sources from Rapid ASKAP Continuum Survey high-band (SPICE-RACS), VLASS, and The HI/OH/Recombination line survey of the Milky Way (THOR).}
\label{tab:survey_specs}
\end{table}

\section{Summary}\label{Conclusion}
In this work, we developed a set of heuristics with which to guide the use of QU-fitting on POSSUM survey data. These rules apply to five commonly used QU spectra models: the Faraday simple (Equation \ref{FS}), Burn slab (Equation \ref{BS}), two-component (Equation \ref{TC}), internal turbulence model (Equation \ref{T-slab}), and external turbulence model (Equation \ref{T-Screen}). For each model, we generated two populations of 10 000 sources each, one for POSSUM low-band and the second with both POSSUM bands. These populations span uniform distributions of model parameters and S:N. We performed QU-fitting to the populations with both their true model and a commonly `confused' model in a series of experiments designed to help to define where in model parameter space QU-fitting confidently selects the true model and where QU-fitting can provide accurate model fits.

From these experiments, we conclude that while all metrics of goodness of fit perform comparably for simple models, only Bayesian evidence (via the Bayes factor) consistently identifies the correct model for complex cases (internal turbulence, external turbulence, and two-component models). We present our guidelines in Table \ref{Conclusion_Table}, for both model selection and limits for which the fits from QU-fitting are reasonably unbiased. 

We also analyzed the reliability of the measurement uncertainties provided by QU-fitting for the model parameters. 
\begin{itemize} 
\item Faraday simple - All the measurement uncertainties are reliable. 
\item Burn slab - All the measurement uncertainties are reliable.
\item Two-Component - None of the measurement uncertainties in either band regime are reliable.
\item Internal Turbulence - The $\sigma_{\text{RM, src, QU}}$ measurement uncertainties are reliable when using the full-band.
\item External Turbulence - The $\text{RM}_{\subtxt{screen, QU}}$ and $\sigma_{\text{RM, FG, QU}}$ measurement uncertainties are reliable when using the full-band. The uncertainties for $p_{0, \text{QU}}$ are accurate in both regimes.
\end{itemize}

We note that our results only consider the scenario of a single, fairly small, band gap. In this work we do not present any results on the effect of large band gaps and/or missing channels on QU-fitting selection or accuracy; we are uncertain of the method's results in these cases. Our work provides a firm basis of the future of QU-fitting with POSSUM. Further analysis done on this method should account for band gaps, removed channels, different surveys, and more advanced models.


\section*{Acknowledgements}
We acknowledge the traditional owners of the land on which the ANU stand, the Gamilaraay, the Ngunnawal and the Ngambri people, as well as the  Wajarri Yamaji people as the Traditional Owners and native title holders of the Inyarrimanha Ilgari Bundara, the Murchison Radio-astronomy Observatory site.

This work made use of the following software packages: \texttt{astropy} \citep{astropy2013, astropy2018, astropy2022}, \texttt{Jupyter} \citep{Perez2007, kluyver2016jupyter}, \texttt{matplotlib} \citep{Hunter2007}, \texttt{numpy} \citep{numpy}, \texttt{pandas} \citep{mckinney-proc-scipy-2010, pandas_16918803}, \texttt{python} \citep{python}, and \texttt{scipy} \citep{2020SciPy-NMeth, scipy_17101542}.
\\
\section*{Data Availability}
The data used in this work was simulated and can be made available to authors upon request.



\bibliographystyle{mnras}

\newpage


\appendix
\section{QU Models Mathematical Degeneracy}
\label{app:Degen_Proof}
In this Appendix, we highlight cases where models representing physically distinct lines of sight can become mathematically degenerate once convolved with the instrument beam. Our work extends that of \citep{Sokoloff1988}, who formalized combining Faraday rotation and wavelength-independent depolarization to create line-of-sight models of complex polarization (as well as \cite{Burn1967} and \cite{Tribble1991}). By assuming a background synchrotron-emitting source bounded at $z = z_{\mathrm{S}}$ along the line of sight, and a thermal plasma bounded at $z = z_{\mathrm{T}}$ with $z_{\mathrm{S}} < z_{\mathrm{T}}$, \cite{Sokoloff1988} derived the resulting complex polarization as the product of the intrinsic synchrotron polarization, $\mathbf{{P}_{\mathrm{int}}}$, and the contribution from external Faraday rotation, $\mathbf{{P}_{\mathrm{ex}}}$:

\begin{equation}
\mathbf{P} = \mathbf{P}_{\mathrm{int}} \mathbf{P}_{\mathrm{ex}}.
\label{original}
\end{equation}

This is the formalism used to create the models tested in this paper (Burn slab, two-component, internal turbulence, and external turbulence). However, not all combinations, especially when convolved with a Gaussian beam, are mathematically distinct. This has implications for the future of line-of-sight physics modeling, specifically as the resolution of instruments grows and the finer polarization structure of a line-of-sight becomes observable. For this example we use the models for a background Faraday simple source Equation (\ref{FS}), combined with a foreground screen with a linear RM gradient when convolved with a Gaussian beam:

\begin{equation}
      e^{2i \lambda^{2} \text{RM(x)}} .
\end{equation}
Where RM(x) = $\text{RM}_{\subtxt{screen}} + \frac{\Delta \text{RM}x}{\sigma_{b}}$, and $\Delta \text{RM}$ is the \text{RM} gradient across the beam, x is the coordinate on the sky (with x = 0 at the source center), and $\sigma_{b}$ is the FWHM of the beam. To calculate the complex polarization we combine the source and foreground model according to Equation \ref{original} and then convolve to a Gaussian beam according to the convolution principle:
\begin{equation}
    (f * g)(\tau) = \int f(\tau - t)g(\tau) d\tau .
\end{equation}
The full model for this complex polarization along the line of sight is:
\begin{equation}\label{full_unf_em_sc_beam_derivation}
     \begin{split}
        \mathbf{P}(\lambda^{2})  = p_{0} e^{2i\psi_{0}}  \int\limits_{-\infty}^{\infty} \int\limits_{-\infty}^{\infty} \frac{1}{2\pi\sigma_{b}} e^{\frac{((x^{\prime}-x)^{2} + ((y^{\prime}-y)^{2}}{2\sigma_{b}^{2}}} e^{2i \lambda^{2} \text{RM} (x^{\prime} - x))} dx^{\prime}dy^{\prime}
     \end{split} .
\end{equation}
After removing constants and rearranging, this integral can be solved by either completing the square or integration by parts. Both methods arrive at the equation: 
\begin{align}
    -\frac{1}{2\sigma_{b}^{2}} \bigg[
    \left( x^{\prime\prime} - 2i\lambda^{2} \frac{\Delta \text{RM}}{D} \sigma_{b} \right)^{2}
    - 4\lambda^{4} \frac{\Delta \text{RM}^{2}}{D^{2}} \sigma_{b}^{2}
    \bigg] .
    \label{eq:factored_exponent}
\end{align}
Complete square and redistribute the $\frac{-1}{2\sigma_{b}^{2}}$:
\begin{equation}
     \frac{-(x^{\prime \prime} - 2i\lambda^{2} \frac{\Delta \text{RM}}{D}\sigma_{b}^{2})^{2}}{2\sigma_{b}^{2}} - \frac{4\lambda^{4} \frac{\Delta \text{RM}^{2}}{D^{2}}  \sigma_{b}^{2}}{2\sigma_{b}^{2}} .
\end{equation}
This integral can be simplified to:
\begin{align}
     \mathbf{P}(\lambda^{2}) = p_{0} e^{2i\psi_{0}} \frac{1}{2\pi\sigma_{b}^{2}} 
    e^{2i\lambda^2 \text{RM}_{0}} \nonumber \sqrt{2} \sqrt{\pi} \sigma_{b} 
    e^{-2 \lambda^{4} \Delta \text{RM}^{2} \sigma_{b}^{2}} \nonumber \\
      \int\limits_{-\infty}^{\infty} 
     e^{\frac{(x^{\prime \prime} + 2i\lambda^{2} \frac{\Delta \text{RM}}{D} \sigma_{b})^{2}}{2\sigma_{b}^{2}}} 
    dx^{\prime \prime},
    \label{eq:my_equation}
\end{align}
a complex Gaussian, the integral of which is evaluates to become $\sqrt{2}\sqrt{\pi}\sigma_{b}$. The $\frac{1}{2\pi\sigma_{b}^{2}}$ term cancels with the two $\sqrt{2}\sqrt{\pi}\sigma_{b}$ terms, so that the result is
\begin{equation}
\mathbf{P}(\lambda^{2})
= p_{0} e^{2i\psi_{0}}
  e^{2i\lambda^2 \mathrm{RM}_{\mathrm{screen}}}
  e^{-2 \lambda^{4}
    \frac{(\Delta \mathrm{RM})^{2}}{D^{2}}
    \sigma_{b}^{2}} .
\label{unsimpl_result}
\end{equation}
Finally, we recognize that the change in the \text{RM} on the scale of the beam, $\frac{\Delta \text{RM}^{2}}{D^{2}} \sigma_{b}^{2}$ is the definition of $\sigma_{\subtxt{RM, FG}}$, thus we rewrite Equation \ref{unsimpl_result} as:
\begin{equation}
\mathbf{P}(\lambda^{2}) = p_{0} \exp{2i(\psi_{0}+\text{RM}_{\subtxt{screen}}\lambda^{2})}  \exp{-2\lambda^{4}\sigma_{\subtxt{RM, FG}}^{2}},  
\end{equation}
or Equation \ref{T-Screen}. This model physically represents a linear gradient shift in rotation measure in the foreground but is mathematically the same as if there was a random foreground when convolved to the Gaussian beam. This is just one of many model combinations that convolve to become degenerate. 

\section{Signal-to-noise}\label{SN}
In this Appendix, we derive the per-channel noise amplitude from the band averaged S:N. We define the band-averaged signal-to-noise ratio as the mean ratio of the polarized fraction ($ S_{\text{band}}$), divided by the variance, ($\langle S_{\text{band}} \rangle$ / $\mathrm{Var(S_{band})}$). For the  models, the average noise-free signal can be calculated as $S_{\text{band}}$ = $\sum^N_{i=1} p_{i} / N$ where $N$ is the number of channels and $p_{i}$ is the polarization of channel $i$. We calculate our band average noise from the signal variance:
\begin{equation}
    \mathrm{Var(S_{band})} = \sum_{i=1}^{N} \mathrm{Var} \left ( \frac{p_{i}}{N} \right ).
\end{equation}
According to the properties of the variance and assuming that all channels have the same noise variance we can rewrite the equation for the band averaged noise to be: 
\begin{equation}
    \mathrm{Var(S_{band})} = \sum_{i=1}^{\mathrm{N}} \frac{1}{\mathrm{N}^{2}}\mathrm{Var(p_{i})} = \frac{1}{\mathrm{N}^{2}} \sum_{i=1}^{\mathrm{N}} \sigma_{S,i}^{2} = \frac{N}{\mathrm{N}^{2}} \sigma_{S,i}^{2} = \frac{\sigma_{S,i}^{2}}{\mathrm{N}},
    \label{noise}
\end{equation}
where $\sigma_{S, i}$ is the standard deviation on the measured polarization for $i$th channel. We define the band averaged S:N as:
\begin{equation}
    \mathrm{S\mathpunct{:}N_{band}} = \frac{\mathrm{\langle S_{band} \rangle}}{\mathrm{Var(S_{band})}} =   \frac{\frac{\sum_{i=1}^{N} p_{i}}{N}}{\frac{\sigma_{S,i}^{2}}{\sqrt{N}}} = \frac{\sum_{i=1}^{N} p_{i}}{\sigma_{S,i}} \frac{1}{\sqrt{N}} 
\end{equation}

Since we have made the $\mathrm{S\mathpunct{:}N_{band}}$ a parameter of the models and assuming that all channels have the same noise amplitude, we can then calculate the noise amplitude for a given source with a signal and $\mathrm{S\mathpunct{:}N_{band}}$ as:

\begin{equation}
    \sigma_{S,i} = \frac{\sum_{i=1}^{N} p_{i}}{\mathrm{S\mathpunct{:}N_{band}} \times \sqrt{N}}.
    \label{noiseamp}
\end{equation}
For each source the realizations of the channel noise for Stokes q and u is then drawn from the normal distribution $\mathcal{N}(0,\sigma_{S,i})$.

\section{Goodness-of-fit metrics and scale}\label{Computation}
In this appendix we provide more details on why we selected the Bayes factor to perform our model selection; and provide empirical support for our choice of the Jeffery's scale that we use for model selection strength classification. 

\subsection{Most useful goodness-of-fit metrics}\label{most}
In this section, we evaluate the performance of four goodness-of-fit metrics—$\chi^{2}$, AIC, BIC, and Bayesian evidence—for QU model selection. We analyze their effectiveness across multiple test cases and recommend best practices for both individual and multi-model fitting. For each simulated population, we calculated the percentage of sources for which each metric correctly identified the true model when compared to an alternate.
\begin{figure*}
\includegraphics[width=0.75\textwidth]{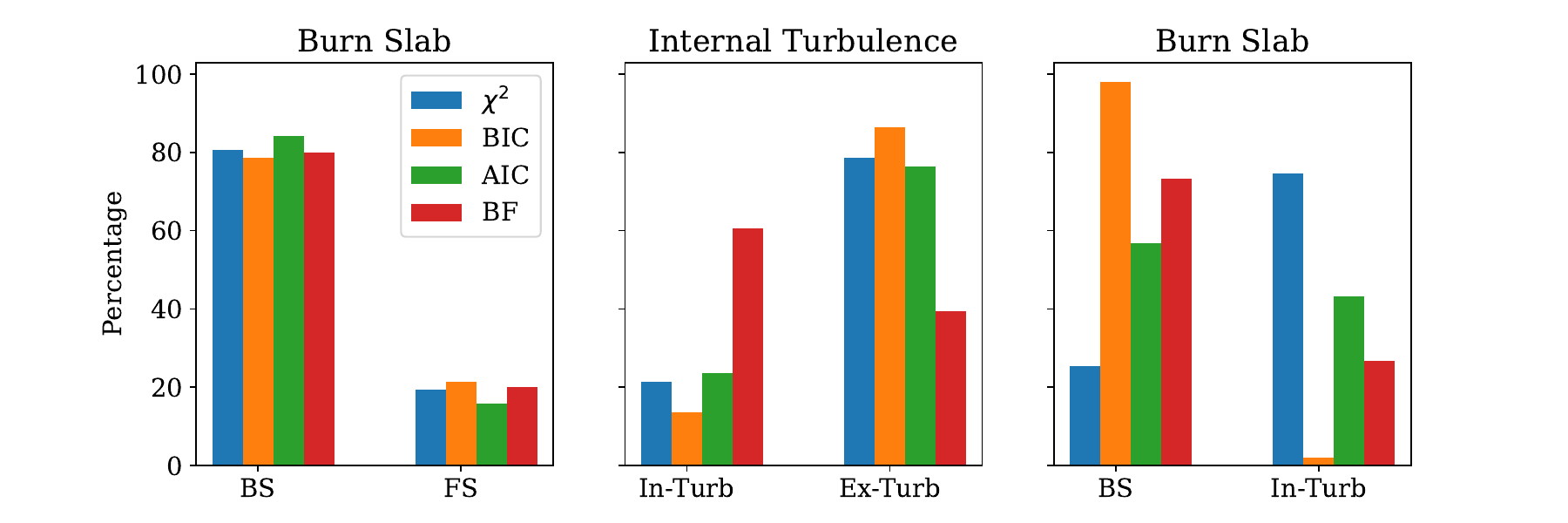}
\caption{The percentage of each model's low-band source population (title) that $\chi^{2}$ (blue), AIC (green), BIC (orange), and Bayesian evidence (red) confidently select the correct model (left) and the alternate model (right). Correct classification thresholds are $\Delta$BIC, $\Delta$AIC, and $\Delta\chi^{2} < -5$, and $\ln(\text{BF}) > 2$.}
\label{fig:bars}
\end{figure*}
Figure \ref{fig:bars} presents results for three key cases. In comparisons involving simple models (e.g., Burn slab vs. Faraday simple), all metrics correctly classifying around 80\% of the population. In contrast, for more complex models such as the internal or external turbulence, only the Bayes factor (derived from Bayesian evidence) consistently performed well, achieving over 80\% accuracy. Other metrics, particularly BIC and AIC, underperformed—often penalizing model complexity even when the complex model was correct. This indicates that these criteria may be overly biased toward simpler models.\\
In tests where the simpler model was indeed correct (e.g., Burn slab vs. internal turbulence), BIC performed best, as expected, while the Bayes factor remained competitive (correctly classifying  \textasciitilde76\%). In these scenarios, $\chi^{2}$ notably overestimated complexity, leading to poor classification.\\
For studies limited to simple models, all metrics—including $\chi^{2}$, AIC, and BIC—are sufficient. However, for broader applications involving both simple and complex models, the Bayes factor is the only metric that performs reliably. It effectively balances fit quality and model complexity and avoids both overfitting and simplicity bias.
In summary, traditional metrics suffice for simple model comparisons, but for confident accurate selections involving complex, multi-parameter models, the Bayes factor is essential. The remainder of our analysis relies on Bayes factor results exclusively.

\subsection{True positive rate}\label{TPR}
To assess the robustness of the classification scale we examine the true positive rate (TPR) across model parameter space for all the experiments. We used uniform distributions of model parameters when we generated our source populations, which ensures meaningful TPR estimates, interpreted as the probability of correct classification at a given parameter value. 

\begin{figure}
	\includegraphics[width=0.5\textwidth]{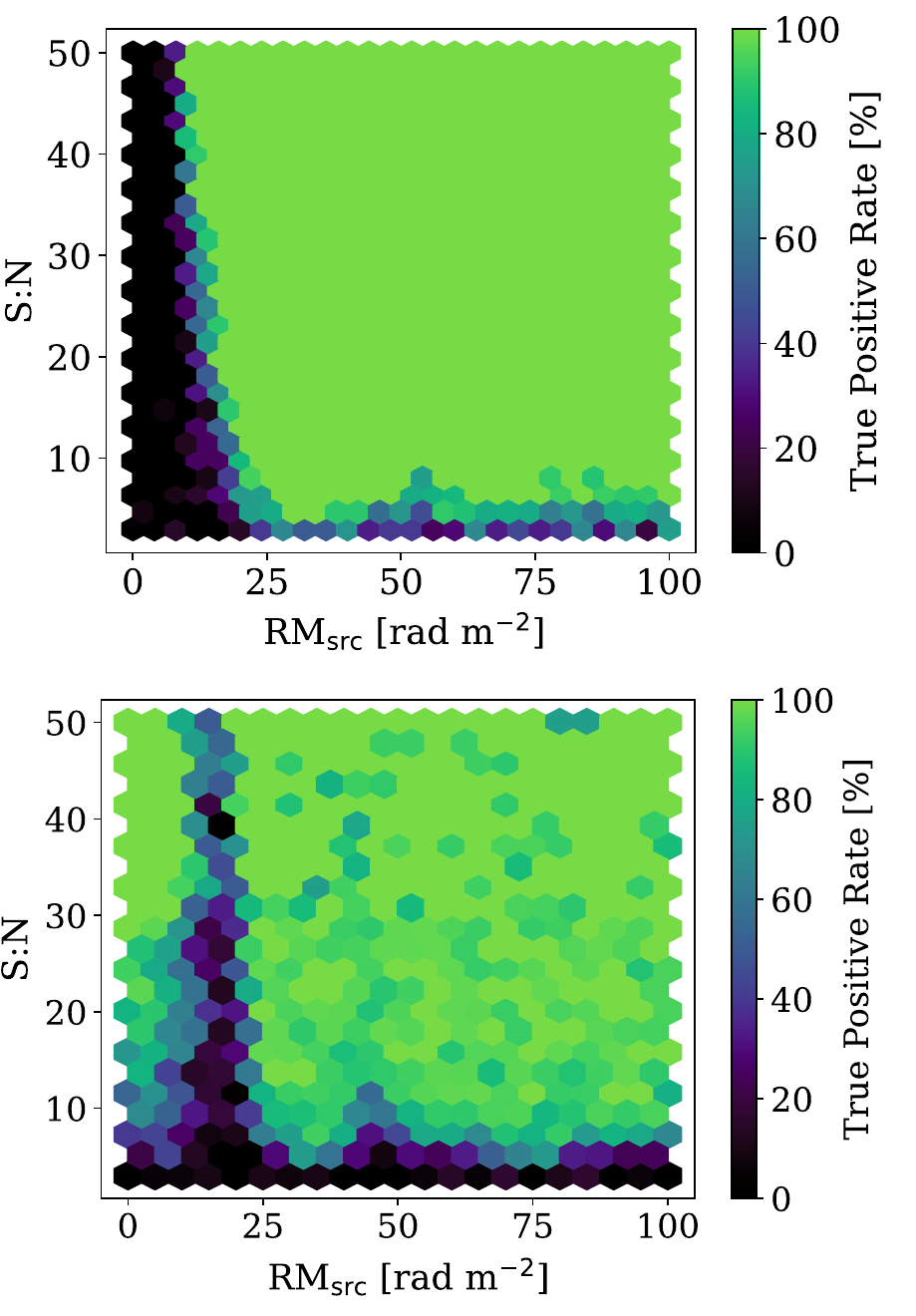}
    \caption{The true positive rates for the slab width and signal-to-noise parameter space for the population of true Burn slabs when fit with the Faraday simple source (left) and the internal turbulence (right) as the alternate models.}
    \label{fig:percentages}
\end{figure}

In Figure \ref{fig:percentages} we present the results for two of the experiments.  We show the S:N -- $\text{RM}_{\subtxt{src}}$ space colored by TPR for two cases, the Burn slab vs Faraday simple model (upper panel) and the Burn slab vs the internal turbulence model (lower panel). These maps show the same parameter space shown in Figures \ref{fig:BF_parameter_space} (upper panel) and \ref{fig:MS_BS_band1} (lower panel).  We find that bins with median Bayes factor classifications above 5.0 also have a TPR $\approx $ 100 \%. While bins with median Bayes factor of 1.0 - 5.0 (for example at S:N $\approx$ 6 and less than the limit defined by Equation \ref{FSBSrelation} for the Burn slab vs. the Faraday simple)  the TPR decreases to 60-50 \%. We find similar trends in the other experiments as well, with the TPR closely aligned with our selection classification scale. These results support our choice to follow \cite{jeffreys1961} strength classification. For the full-band, the additional bandwidth enables larger fractions of confidence classification at lower S:N and for thinner slabs compared to the low-band case (see Appendix \ref{app:AccuracyPlots} for fits)

\section{Full-band QU Selection Results}
\label{app:AccuracyPlots}
\begin{figure*}
	\includegraphics[width=0.750\textwidth]{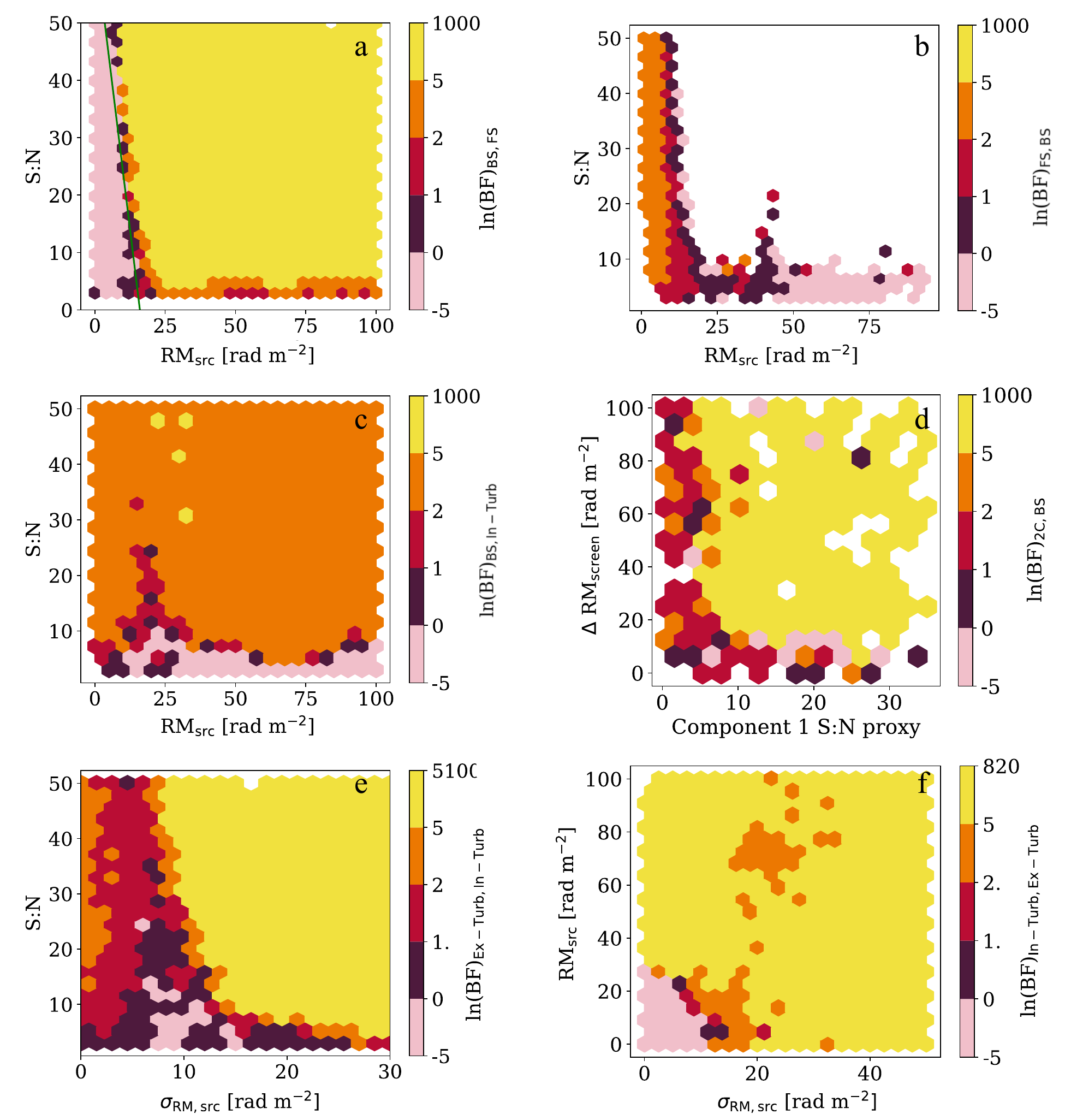}
    \caption{This figure contains the full-band (the low-band 800-1088 MHz and the mid-band 1296-1440 MHz) parameter maps for the model selection between: a) the true Burn slab population and Faraday simple, b) the true Burn slab and the internal turbulence, c) the true external turbulence  and the internal turbulence, d) the true Faraday simple and the Burn slab, e) the true two component population and the Burn slab model, and f) the true internal turbulence and the external turbulence. They are binned using the scale shown in Table \ref{tab:binning}}
    \label{fig:combined_band_selection}
\end{figure*}

This section contains the full-band population results for the parameter space selection maps. Figure \ref{fig:combined_band_selection} contains the full-band results for sections \ref{Models_Exp_Results}. In most cases except the two-component model the additional bandwidth does improve the outcome of the model selection. 
\begin{figure*}
	\includegraphics[width=0.60\textwidth]{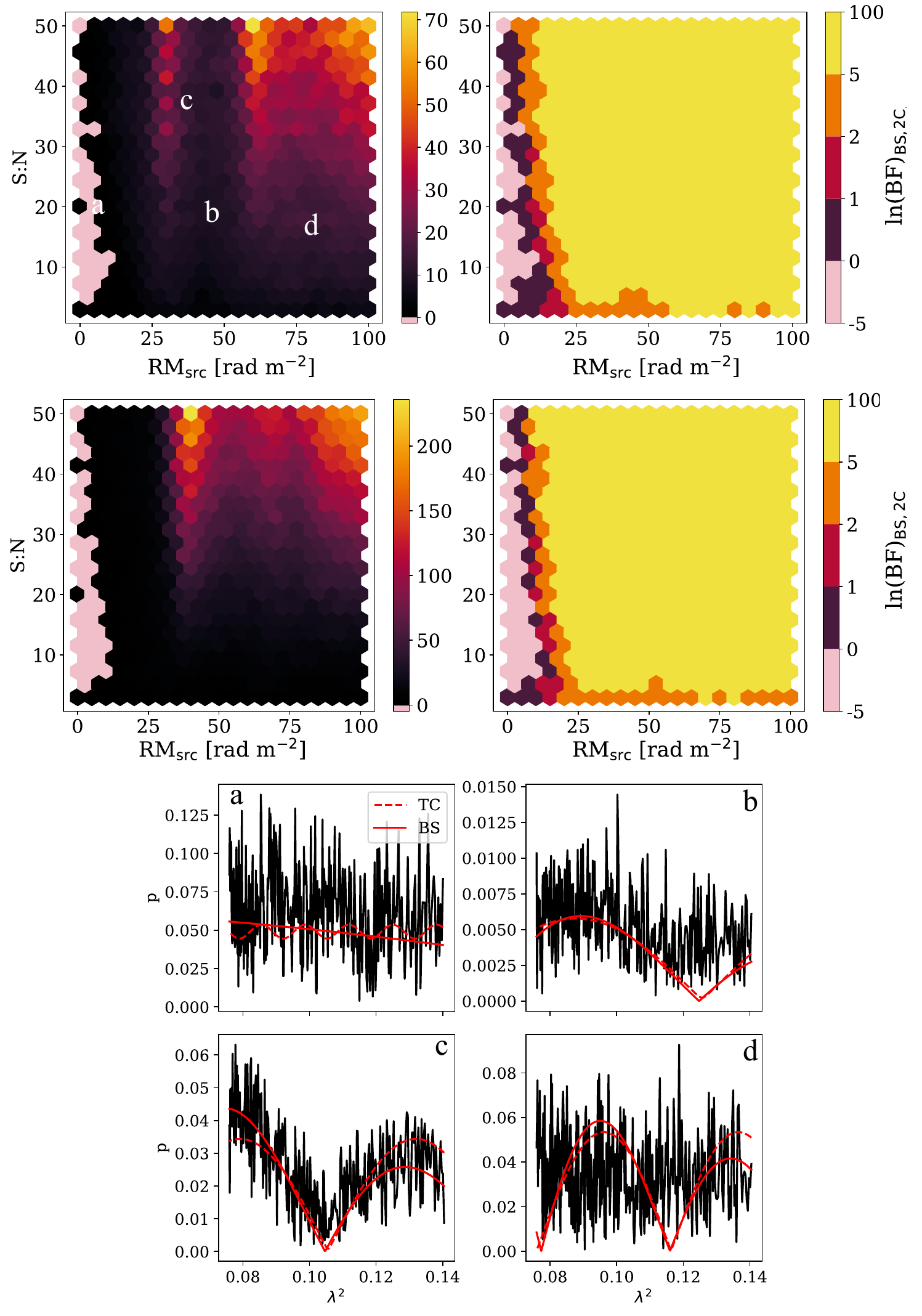}
    \caption{The parameter selection map for the population of true Burn slabs when fit with the two component model as the alternative. The populations are binned by slab width ($\text{RM}_{\subtxt{src}}$) and S:N. We present both the ln(BF) numbers and apply the confidence bins shown in Table \ref{tab:binning}. The four spectra corresponds to the labels positions on the a parameter space maps. The dashed and solid lines refer to the best fit from QU-fitting for the two component and Burn slab models respectively.}
    \label{fig:BS_TC_fullselection}
\end{figure*}

\section{QU Model Atlas}
\label{app:Atlas}
Within this section we present a QU spectra atlas for all models discussed in this paper (Burn slab, two-component,internal turbulence model and random foreground screen respectively). 

\begin{figure*}
    \includegraphics[width=0.95\textwidth]{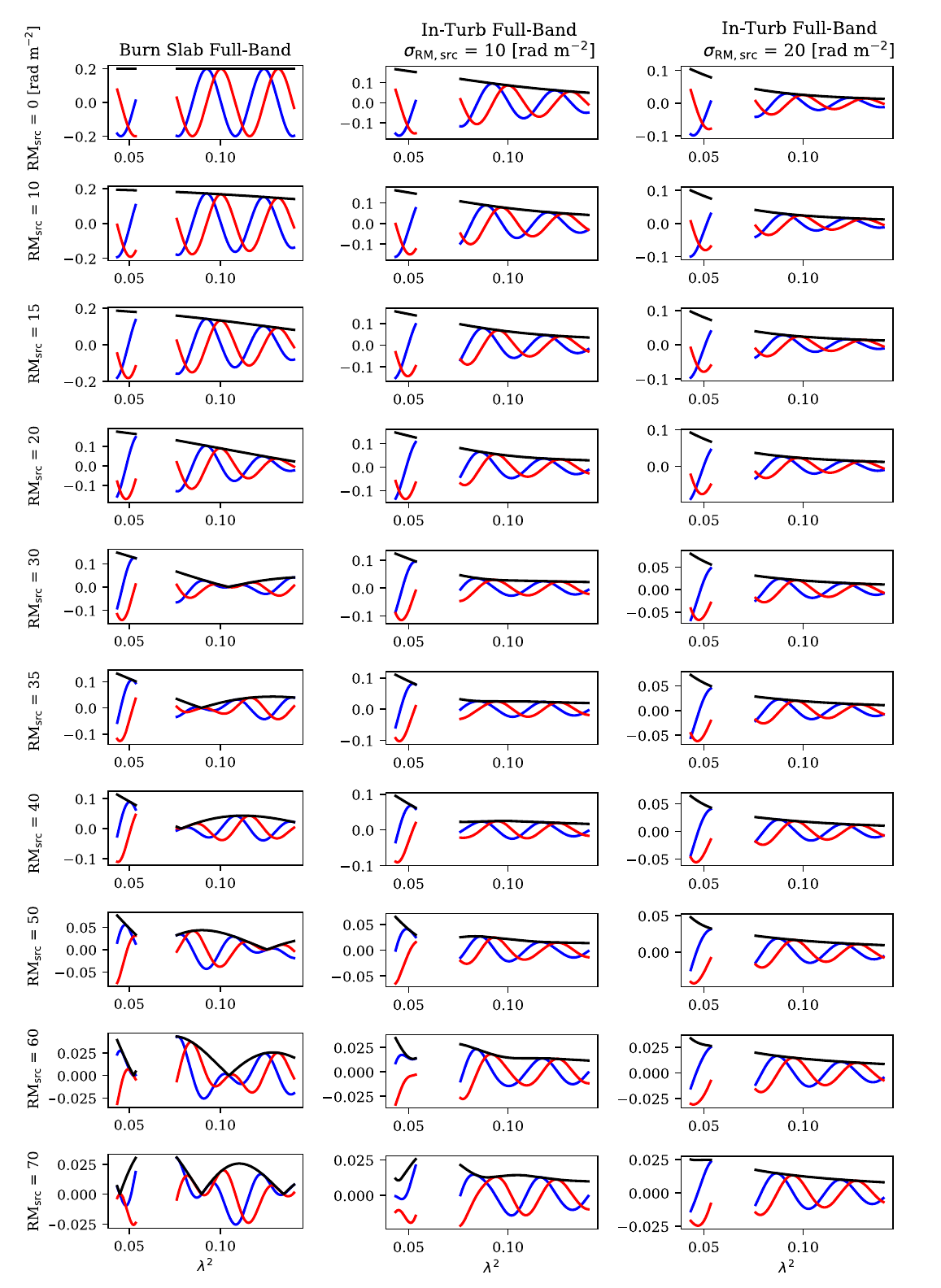}
    \caption{An atlas of full-band (low-band inclusive) qu-spectra for the Burn slab model and the internal turbulence model. The first row is the Burn slab spectra with in increasing $\text{RM}_{\subtxt{src}}$ (top to bottom). The last two rows are the spectra for the internal turbulence model for two different $\sigma_{\subtxt{RM,src}}$. We note the 'bump' feature present in the spectra at  $\sigma_{\subtxt{RM,src}} =  10$ rad m$^-2$. This feature is the key to QU-fitting being able to correctly estimate the model parameters.} 
    \label{dict_BS_MS}
\end{figure*}

\begin{figure*}
    \includegraphics[width=0.65\textwidth]{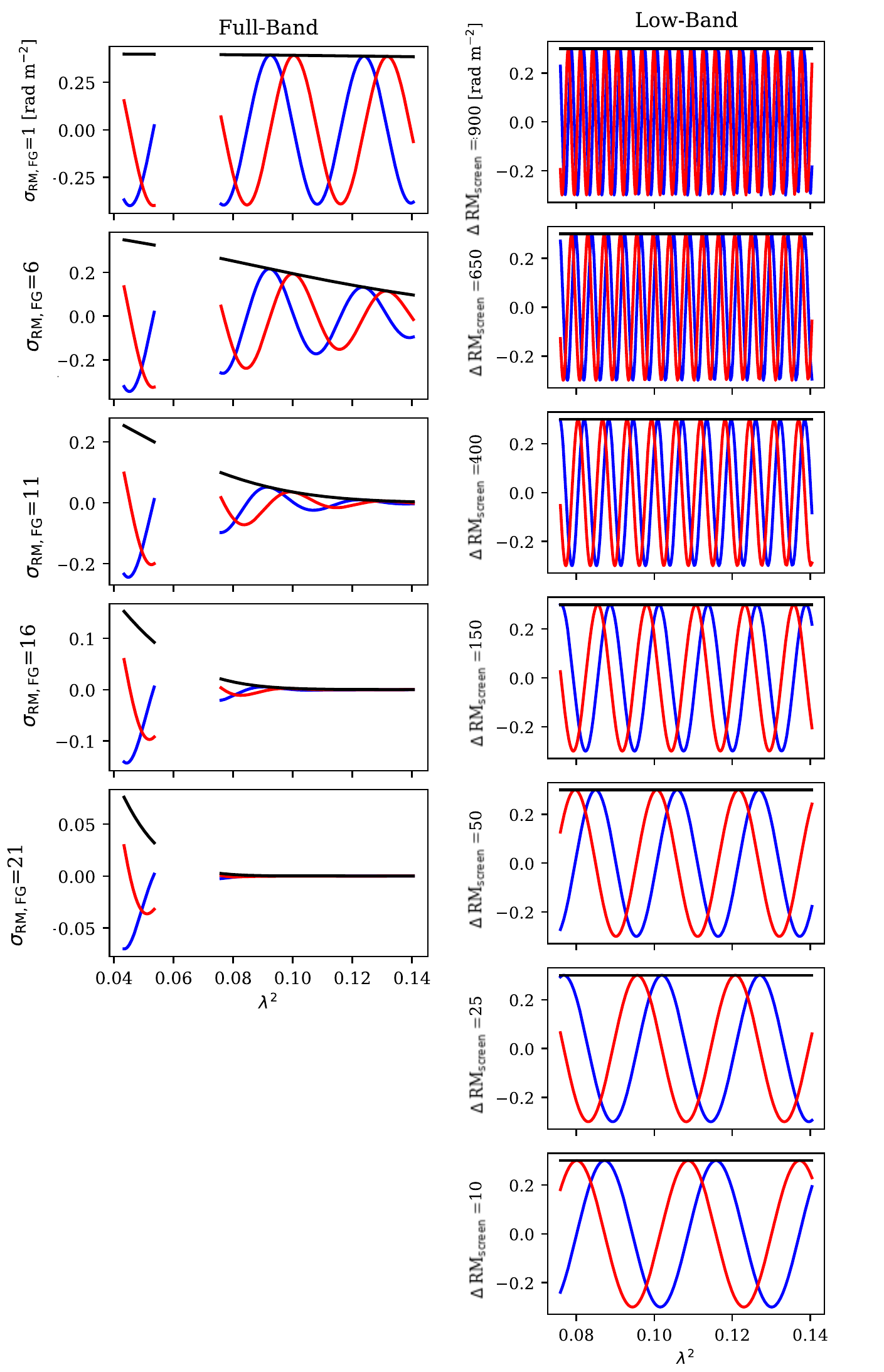}
    \caption{An atlas of $QU$-spectra for the full-band external turbulence model (first column) and the two-component model (second column). The spectra are in order (top to bottom) of increasing $\sigma_{\subtxt{RM, FG}} $ and decreasing $\text{RM}_{\subtxt{screen}}$ separation for the second column.} 
    \label{dict_RFG}
\end{figure*}

In Figure \ref{dict_BS_MS} we present both the low-band and full-band spectra for the Burn slab and internal turbulence models. These spectra are, from top to bottom, in order from increasing $\text{RM}_{\subtxt{src}}$. In Figure \ref{dict_RFG} we show the evolution of the external turbulence spectra with increasing $\sigma_{\text{RM,FG}}$ for both the low-band and full band in the first two columns. The third column shows the two-component spectra with increasing separation (top to bottom) between the components. 

\begin{figure*}
    \includegraphics[width=0.95\textwidth]{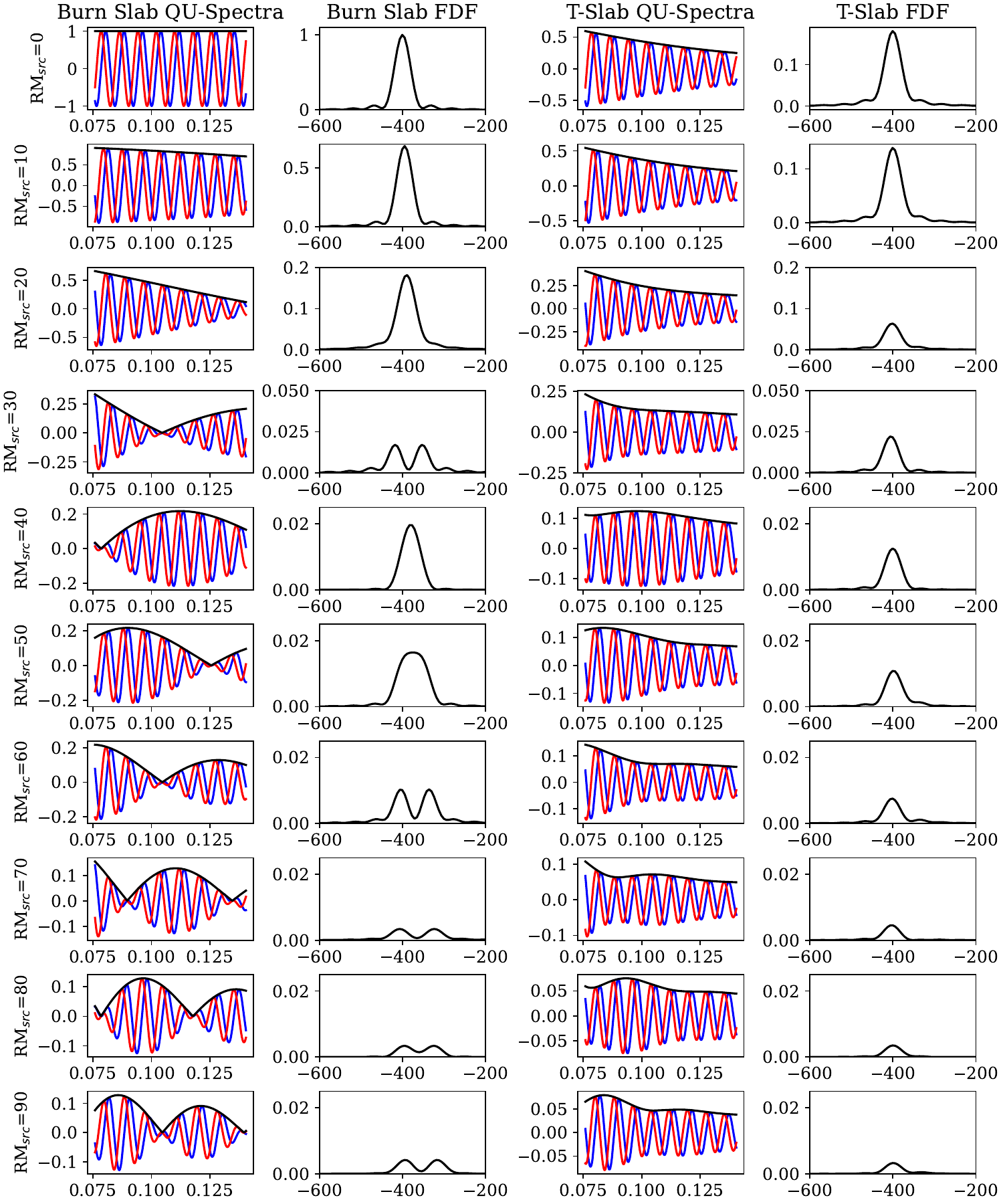}
    \caption{An atlas of $QU$-spectra and corresponding Faraday depth space for the Burn slab and the internal turbulence model for the POSSUM low-band. The slab width $\text{RM}_{\subtxt{src}}$ increases from top to bottom. We note that in Faraday depth space it can be difficult to separate the regular field component from the effects of both a regular field and small scale field. There is potential with QU-fitting to differentiate these two models.} 
    \label{dict_BS_TS}
\end{figure*}

In Figure \ref{dict_BS_TS} we show the spectra and the Faraday Depth Function (FDF) for the Burn slab and the internal turbulence model. We note that although these two models can appear similar in FDF space they are perhaps more distinct in QU space.

\bsp	
\label{lastpage}
\end{document}